\documentclass{aa}  

\usepackage{graphicx}
\usepackage{txfonts}
\usepackage{siunitx}
\usepackage{hyperref}
\usepackage{tikz}

\usepackage{xcolor}

\usepackage{subfigure}
\usepackage{soul}
\setstcolor{red}

\begin{document}

   \title{Bridging Simulations and Observations: New Insights into Galaxy Formation Simulations via Out-of-Distribution Detection and Bayesian Model Comparison}
   \titlerunning{Evaluating Cosmological Simulations using Machine Learning Tools}
   \subtitle{Evaluating Galaxy Formation Simulations under Limited Compute Budgets and Sparse Dataset Sizes}

   \author{Lingyi Zhou
          \inst{1}
          \and
          Stefan T.~Radev
          \inst{3}
          \and 
          William H. Oliver
          \inst{1,2}
          \and 
          Aura Obreja
          \inst{1,2}
          \and
          Zehao Jin
          \inst{4}
          \and
          Tobias Buck
          \inst{1,2}
          }

   \institute{Interdisciplinary Center for Scientific Computing (IWR), University of Heidelberg,
 Im Neuenheimer Feld 205, D-69120 Heidelberg, Germany\\
 \email{lingyi.zhou98@outlook.com}
 \and
 Universität Heidelberg, Zentrum für Astronomie, Institut für Theoretische Astrophysik, Albert-Ueberle-Straße 2, D-69120 Heidelberg, Germany\\
 \email{tobias.buck@iwr.uni-heidelberg.de}
 \and 
 Center for Modeling, Simulation, \& Imaging in Medicine, Rensselaer Polytechnic Institute, NY, USA
 \and 
 Center for Astrophysics and Space Science (CASS), New York University Abu Dhabi \\
             }

   \date{Received Month, XXXX; accepted Month Day, 7/7/2025}

 \abstract
   {Cosmological simulations are a powerful tool to advance our understanding of galaxy formation. A question that naturally arises in light of high-quality observational data is: How close are the models to reality? Due to the high-dimensionality of the problem, many previous studies evaluate galaxy simulations using simplified summary statistics.}
   {In this work, we combine simulation-based Bayesian model comparison with a novel misspecification detection technique to compare galaxy images of 6 hydrodynamical models from the NIHAO and IllustrisTNG simulations against observations from SDSS.}
   {Since cosmological simulations are computationally costly, we first train a $k$-sparse variational autoencoder (VAE) on the abundant dataset of SDSS images. The VAE learns to extract informative latent embeddings and delineates the typical set of real images. To reveal simulation gaps, we perform out-of-distribution (OOD) detection based on the logit functions of classifiers trained on the embeddings of simulated images. Finally, we perform amortized Bayesian model comparison using probabilistic classification, identifying the relatively best-performing model along with partial explanations through SHAP-values.}
   {We find that all 6 models are misspecified compared to SDSS observations and can only explain part of reality. The relatively best performing model comes from the standard NIHAO simulations without AGN physics. Carefully inspecting SHAP-values we find that the main difference between NIHAO and IllustrisTNG is given by color and morphology -- NIHAO is redder and clumpier than IllustrisTNG.}
   {By using explainable AI methods such as SHAP values in combination with innovative methods from simulation-based Bayesian model comparison and new misspecification detection techniques we are able to quantitatively compare costly hydrodynamical simulations with real observations and gain physical intuition about the quality of the simulation models. Hence, our new methods help explaining which physical aspects of a particular simulation causes it to match real observations better or worse. This unique feature helps to inform simulators on how to improve their simulation model. }

\maketitle
\section{Introduction}
\label{sec:intro}

Investigating the physical processes that govern the formation and evolution of galaxies is a hard problem. Many of these processes, which span a very large dynamical range, are coupled, and thus, understanding their importance for galaxy formation requires running cosmological hydrodynamical simulations \citep{vogelsberger2020cosmological}. However, assessing the quality and realism of these simulations is a notoriously difficult task. A common approach is to compare the distribution of galaxy properties retrieved from simulations and observations as a diagnostic tool. However, galaxy observations span a multi-dimensional, complex parameter space (image-like or time series-like data), and it is not clear how to optimally perform model comparison in such a setup.
Many previous works have measured the gap between simulation models and observations using traditional methods employing simple 2d or 3d summary statistics. For example, the Tully-Fisher relation between the luminosity of a spiral galaxy and its rotation velocity \citep{tully1977new}, the joint distribution of luminosity, optical rotation velocity and disk size of spiral galaxies \citep{courteau2007scaling}, and the stellar mass and halo mass relation \citep{Moster2013,Moster2018}. Many modern studies try to match multiple observed properties of galaxies. For instance, UniverseMachine \citep{behroozi2019universemachine}, a state-of-the-art algorithm for predicting observable galaxy properties based on simulations, is optimized to simultaneously match a wide range of these properties. However, this is a very limited criterion, as a model may closely match real observations under one such relation, but deviate significantly from reality under another. 

Even worse, a significant challenge arises when attempting to quantitatively evaluate the importance of different sets of summary statistics in the face of contradictions. In such cases, it becomes essential to explore alternative methods for performing comparison, maximizing the potential of the high resolution achieved by simulations and observations.

A natural choice here are galaxy images. Compared to simple summary statistics with high information loss, galaxy images contain a wealth of detailed information. Many previous works compare the distribution of image-based parameters in mock images with that in real observations using a variety of metrics and statistical tools, which have become a key tool in calibrating modern simulations \citep{snyder2015galaxy, bottrell2017galaxiesA, bottrell2017galaxiesB, rodriguez2019optical, bignone2020non, de2022observed}. These include parametric methods like the S{\'e}rsic parameters \citep{sersic1963bol}, as well as non-parametric ones such as concentration, asymmetry, clumpiness statistics \citep[CAS][]{conselice2003relationship}, Gini-$M_{20}$ method \citep{lotz2004new} for identifying whether a galaxy has experienced a recent merger event and the multimode, intensity and deviation statistics \citep[MID][]{freeman2013new}.

The rapid development of artificial intelligence (AI) has had a profound impact in many fields, including astrophysics. In particular, it improves our understanding of images as well as the comparison of simulations and real observations. Numerous studies have applied machine learning methods to analyze galaxy images, thereby addressing challenges in astrophysics \citep{dieleman2015rotation, obreja2018introducing, buck2021predicting, buder2021galah+, cheng2021beyond, storey2021anomaly, smith2022realistic, tohill2024robust}.

Recently, several works have explored machine learning (ML) approaches to compare simulations and observations. \citet{karchev2023simsims} used the deep learning method for hierarchical models proposed by \citet{elsemüller2023deep} to compare simulation-based supernova Ia light curve models. \citet{schosser2024optimal} used a 3D-CNN as the summary network compressing 3D light cone data to 6-dimensional latent vectors and an invertible neural network conditioned on an observation \citep[cINN ][]{ardizzone2018analyzing} as the inference network for extracting model parameters posterior. \citet{zanisi2021deep} compared Illustris \citep{vogelsberger2014introducing} and IllustrisTNG \citep{pillepich2018simulating} with $r$-band Sloan Digital Sky Survey \citep[SDSS][]{kollmeier2019sdss} images by combining the output of two PixelCNN networks \citep{van2016pixel} to produce pixel-wise anomaly scores assigned to simulation images. \citet{jin2024quantitatively} proposed to use GANomaly \citep{akcay2019ganomaly}, an anomaly detection network based on Generative Adversarial Networks \citep[GAN][]{goodfellow2020generative}, to rate NIHAO simulations \citep[Numerical Investigation of Hundred Astrophysical Objects][]{wang2015nihao, buck2019nihao, buck2020nihao} against SDSS images by assigning anomaly scores to galaxy images. 
In addition, \cite{2020Margalef-Bentabol} employed Wasserstein generative adversarial networks (WGANs) \citep{wgan} to find outliers in Horizon-AGN simulation \citep{Dubois2014} using H-band CANDELS \citep{Grogin2011,Koekemoer2011} images and the WGAN loss as anomaly score. In another approach, \citet{Eisert2024} used representation learning techniques, specifically contrastive learning, to encode mock IlustrisTNG images and real HSC images into a joint embedding space. Among others, they use this embedding space in combination with nearest neighbour search to identify outlier galaxies.

A fundamental problem of all these approaches is that they rely on large training sets. However, galaxy formation simulations are computationally 
expensive ($\sim10-100$k CPUh per instance), so we propose a novel approach here. We leverage a large set of real images (643,553) to pre-train a sparse embedding network which compresses simulated and real galaxy images into a structured latent space, which allows us to highlight notable simulation gaps \citep{schmitt2023detecting}. Then, we use the amortized Bayesian model comparison (BMC) \citep{radev2021amortized, radev2023bayesflow}, which is a novel simulation-based inference \citep[SBI;][]{cranmer2020frontier} method for comparing analytically intractable, high-dimensional models, to determine the relative fit of each model. This allows us to efficiently handle a large number of images, which would be computationally infeasible with standard Bayesian methods. Given the limited size of our simulation dataset -- an inherent challenge for SBI applications that typically require large amounts of data -- we utilize ensemble methods to enhance classifier performance and robustness despite the data scarcity.

The aim of this work is to provide a unified framework to robustly compare different numerical models of galaxy formation against observations. In particular, we aim at comparing models of different parameter choices, different numerical recipes and sample sizes on an equal footing. In contrast to previous methods, our approach begins with a k‑sparse variational autoencoder that is pre-trained on a large set of SDSS images. By learning a latent embedding that is rooted in abundant observational data, our method produces a representation space that better reflects the true distribution of galaxy properties. The use of sparsity in the autoencoder encourages the model to capture the most salient and physically interpretable features, thereby disentangling different aspects of the galaxies in a way that facilitates later analysis. Another approach to find useful representations was recently presented by \citet{Eisert2024} who present a framework that uses a self‐supervised contrastive learning method to map both simulated and observed galaxy images into a 256-dimensional representation space. Their approach makes use of an E(2)-equivariant ResNet with steerable convolutional layers so that the resulting features are invariant to rotations and reflections, ensuring robustness against common observational variations. Their work focuses on aligning simulated and observed distributions and evaluating the overlap in the feature space, mostly in qualitative terms. Our approach, however, takes a further step by casting model comparison as an amortized Bayesian inference problem. We interpret the classification task as deriving probabilistic model posteriors and use calibrated ensemble classifiers to produce these relative probabilities. This framework not only quantitatively ranks the performance of different simulation models but also allows us to overcome the challenge of scarce simulation budgets by leveraging the large, readily available observational dataset.

In terms of outlier detection, \citet{Eisert2024} focus on comparing the distribution of images through visualizations like UMAP projections and nearest neighbour distances, while we have developed a more robust, quantitative approach for out-of-distribution (OOD) detection. \citet{Eisert2024} rely primarily on geometric measures in the learned representation space to detect out-of-domain images. In contrast, here we train an ensemble of classifiers on the latent embeddings generated from simulation images to calculate a Generalized ENtropy (GEN) score \citep{liu2023gen}, which provides a clear, statistical criterion for identifying images that do not belong to the simulation domain. By comparing the GEN score distributions of the simulated data with those of the SDSS test set, we can discard OOD observations before performing model comparison, ensuring that only data compatible with at least one model are used.

A major innovation in our method is the integration of explainable AI tools, specifically SHAP values, which we use to interpret the latent features that drive the model comparison. This analysis links abstract latent dimensions back to concrete physical properties such as color and light concentration. In our results, we observe that certain latent features strongly correlate with these physical characteristics, enabling us to explain why, some simulation model variants appear relatively better than other models. This insight offers actionable guidance for simulation improvement by pinpointing the areas where subgrid physics may need refinement.

Compared to previous model comparison works in galaxy formation, our approach is distinguished by 1) its grounding in a richly detailed observational dataset; 2) its rigorous and quantitative OOD detection mechanism; 3) the application of a Bayesian model comparison framework that produces calibrated and interpretable posterior probabilities; and 4) the use of explainable AI to connect latent features with physical galaxy properties. These advancements not only provide a more robust comparison between simulations and observations but also offer a clear pathway for improving the physical realism of cosmological simulations.

This paper is structured as follows. In \autoref{sec:datasets} we start by describing the datasets of SDSS galaxy images together with the simulated galaxy images followed by \autoref{sec:methods} where we discuss our methods including our $k$-sparse VAE architecture, out-of-distribution detection for model misspecification and our amortized Bayesian model comparison procedure. In \autoref{sec:results} we present our results of out-of-distribution detection, model comparison results and our physical interpretation. Finally, \autoref{sec:conclusions} presents our conclusions and outlook. In \autoref{sec:appendix_code}, we provide the link to our Github repository for reproducibility of our results, and in \autoref{sec:appendix_shap_plots}, we present additional figures of various simulation models.

\section{Image datasets and simulation details}
\label{sec:datasets}

\subsection{Observed galaxy images}
\label{subsec:sdss_images}

The Sloan Digital Sky Survey \citep[SDSS][]{kollmeier2019sdss, castander1998sdss} is one of the most influential astronomical surveys ever conducted. Its main goal is to create detailed multi-dimensional maps of the universe, capturing images and spectra of millions of celestial objects. SDSS operates by capturing high-resolution images and acquiring spectral data that provide important information about the distances, composition, and motion of galaxies, quasars, and stars.

\begin{figure*}
   \centering
   \subfigure[SDSS 100128]{
       \includegraphics[width=0.45\columnwidth]{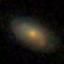}
   }
   \subfigure[TNG100]{
       \includegraphics[width=0.45\columnwidth]{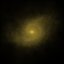}
   }
   \subfigure[AGN]{
       \includegraphics[width=0.45\columnwidth]{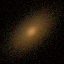}
   }
   \subfigure[NOAGN]{
      \includegraphics[width=0.45\columnwidth]{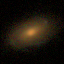}
   }
   \subfigure[SDSS 557508]{
      \includegraphics[width=0.45\columnwidth]{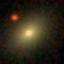}
   }
   \subfigure[TNG50]{
      \includegraphics[width=0.45\columnwidth]{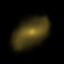}
   }
   \subfigure[UHD]{
      \includegraphics[width=0.45\columnwidth]{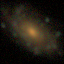}
   }
   \subfigure[n80]{
      \includegraphics[width=0.45\columnwidth]{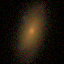}
   }
   \caption{Example rgb images of SDSS (left), TNG100/TNG50 (second from left) and the various NIHAO models of the galaxy g8.26e11 with varying viewing angles (right two columns).
   }
   \label{fig:example-imgs}
\end{figure*}

SDSS images are produced in a set of broad-band filters, of which we use three: the near-infrared ($i$), red ($r$), and green ($g$) which can then be combined into multi-color images by mapping the $i, r, g$-bands to red, green and blue color channels. Following the work of \citet{jin2024quantitatively}, we use the galaxy catalog by \citet{meert2015catalogue} with redshift from 0.005 to 0.395 (with a mean of 0.109), which provides coordinates and stellar masses for 670,722 observed galaxies. To avoid star contamination, we only use galaxies with a stellar mass greater than $10^9M_\odot$ which leaves us with 643,553 galaxy images for the SDSS dataset in our work. SDSS images are cropped to a resolution of 64 × 64 pixels around each galaxy’s coordinates, with the size of the cropped region determined by the pixel scale of the SDSS camera (0.396 arcseconds per pixel). For training our embedding network, we split the SDSS dataset into a training set (70\%), a test set (10\%) and two validation sets: one for early stopping (10\%) and one for hyperparameter tuning (10\%). Two examples of the SDSS images are provided in the left most column of \autoref{fig:example-imgs}.

\subsection{Simulated galaxy images}
\label{subsec:simulated_images}

We compare simulated galaxy images from six different candidate models taken from two different simulation projects: TNG50 and TNG100 from the IllustrisTNG simulations \citep{nelson2019illustristng, pillepich2018first, springel2018first, pillepich2019first, nelson2019first} which only differ in terms of resolution, and four models from the NIHAO simulation suite that vary the physics implementation and resolution: NIHAO-AGN \citep{blank2019nihao, waterval2022nihao}, NIHAO-NOAGN \citep{wang2015nihao}, NIHAO-UHD \citep[Ultra High Definition][]{buck2020nihao} and NIHAO-n80 \citep{Buck2019,maccio2022using}. Below we describe in detail the different aspects of the simulation models.

\subsubsection{IllustrisTNG models}
IllustrisTNG \citep{pillepich2018first,nelson2018first} is a suite of magneto-hydrodynamical cosmological simulations that model the formation and evolution of galaxies within the $\Lambda$CDM cosmology simulated with the moving mesh code Arepo \citep{Springel2010}. The IllustrisTNG suite models galaxy formation in three uniform mass resolution cosmological volume simulations with side lengths $35 h^{-1} \approx 50$ Mpc, $75h^{-1} \approx 100$ Mpc and $205h^{-1} \approx 300$ Mpc, referred to as TNG50, TNG100 and TNG300. In this work we only use data from the former two runs, TNG50 and TNG100.
The TNG models are in particular well suited for studying large-scale structures and statistical properties of galaxies. It offers valuable data on galaxy formation and evolution, the distribution of dark matter, and how explosions from stars and black holes affect galaxies. The only difference between TNG100 and TNG50 is the physical resolution of the underlying simulation.
Sample images from the TNG100 and TNG50 simulations are shown in the second column of \autoref{fig:example-imgs}.

\subsubsection{NIHAO models}
The NIHAO (Numerical Investigation of Hundred Astrophysical Objects) simulation \citep{wang2015nihao} is a suite of hydrodynamical cosmological zoom-in simulations computed with the \textsc{\small GASOLINE2} code \citep{Wadsley2017}. NIHAO adopts a flat $\Lambda$CDM cosmology and parameters from the Planck satellite results \citep{planck}. It includes Compton cooling, photoionization from the ultraviolet background following \cite{Haardt2012}, star formation and feedback from supernovae \citep{Stinson2006} and massive stars \citep{Stinson2013}, metal cooling, and chemical enrichment. A series of prior work has proven that NIHAO simulated galaxies reproduce galaxy scaling relations very well, including the Stellar Halo-Mass relation \citep{wang2015nihao}, the disc gas mass and disc size relation \citep{maccio2016nihao}, the Tully-Fisher relation \citep{dutton2017nihao}, the diversity of galaxy rotation curves \citep{santos2018nihao}, and the mass-metallicity relation \citep{buck2021challenge}. 

In what follows, we refer to this basic version of NIHAO without AGN feedback as `NIHAO NoAGN'. NIHAO NoAGN is the basis for other variations of NIHAO that explore different physical models for star formation and feedback in addition to increased resolution. We describe those versions below.

\begin{itemize}
    \item NOAGN: NOAGN is the vanilla version of the NIHAO simulations as described above.
    \item AGN: The NIAHO model with AGN feedback \citep{blank2019nihao, waterval2022nihao} has the same initial conditions, parameters, and physics as the NOAGN model. It adds Active Galactic Nuclei (AGN) physics in addition to the fiducial physics modeled in NOAGN.
    \item UHD: NIHAO-UHD \citep[Ultra High Definition][]{buck2020nihao} includes a higher resolution version of several Milky Way-like galaxies from the NOAGN model. It has the same initial conditions, parameters, and physics as NOAGN. NIHAO-UHD has produced results that closely match the observed properties of both the Milky Way (MW) and Andromeda (M31) galaxies, such as satellite mass function and MW bulge properties. For more details see \citet{buck2019nihao,buck2018stars}, and \citet{buck2019stars}.
    \item n80: Star formation is typically simulated using a density threshold $n$, measured in particles per ${cm}^{3}$. The transformation of gas particles into star particles begins only when this threshold is attained. All other NIHAO models have a threshold of $n = 10 {cm}^{-3}$ while the n80 model uses a higher value of $n = 80 {cm}^{-3}$ to re-simulate a few galaxies from NIHAO NOAGN \citep{maccio2022using}. The detailed impact of the star formation threshold for the NIHAO simulation is discussed in \citet{dutton2019nihao, buck2019observational} and \citet{dutton2020nihao}.
\end{itemize}

Example images generated from the NIHAO models are provided in the two right most columns of \autoref{fig:example-imgs}.

\subsubsection{Mock observation image pipeline}
All simulated galaxy images are created with the same image pipeline based on radiative transfer (RT) post-processing of the simulated galaxies using the SKIRT code \citep{camps2015skirt}. 
For the IllustrisTNG models we use the synthetic image data from \citet{rodriguez2019optical}, available on the open data access website \footnote{\url{https://www.tng-project.org/data/docs/specifications/\#sec5l}}. We use the redshift zero snapshot (snap number 99) for both TNG100 and TNG50 data. IllustrisTNG galaxies are restricted to subhalos with stellar mass greater than $10^{9.5}M_\odot$ for which morphological measurements are reliable with parameter $flag = 0$ and $sn\_per\_pixel > 2.5$.
For the NIHAO models we use the same synthetic data as \citet{jin2024quantitatively} based on RT postprocessing done by \citet{Faucher2023}. 

For both simulation projects, we create RGB images from the raw RT output following the image pipeline of \citet{jin2024quantitatively} which combines the $i,r,g$ images using an \texttt{arcsinh} stretch as proposed by \citet{lupton2004preparing}, applies a point spread function (PSF) and adds shot noise and Gaussian sky noise to model observational uncertainties following procedures taken from the \textsc{\small RealSim} code by \citet{Bottrell_2019}. More specifically, to account for the SDSS point spread function (PSF), we use a Gaussian PSF with a full width at half maximum corresponding to the average seeing of all SDSS Legacy galaxies: 1.286, 1.356, and 1.496 arcseconds for the $i$, $r$, and $g$ bands, respectively. The physical sizes of the simulated galaxies are converted to angular scales by placing them at a hypothetical redshift of 0.109, matching the mean redshift of our SDSS training sample. Shot noise is modeled as Poisson noise, determined by the survey field parameters, including zero-points, airmass, extinction, and CCD gain. Gaussian sky noise is derived from the average sky noise across all Legacy galaxies. Finally, an \texttt{arcsinh} stretch, as proposed by \citet{lupton2004preparing}, is applied to align with the standard SDSS imaging scheme.

In the following figures, NIHAO models are named with an additional suffix ``rt'' to distinguish them from a previous version of our image data. Since simulated images have different resolutions other than SDSS, we upsample or downsample them to a resolution of 64×64 pixels. Our final simulated image dataset includes 11,334, 1,523, 1,521, 1,540, 120, and 240 images for TNG100, TNG50, AGN, NOAGN, UHD, and n80, respectively. Since this leads to an imbalanced final dataset (TNG100 is the majority class), we oversample the images in the minority classes, see \autoref{subsec:oversampling}. 

For the subsequent steps of model comparison, we stratify the dataset of simulation models into a training set (85\%) and a test set (15\%), ensuring that the class proportions remain unchanged after the split. Importantly, we perform the train-test split \textit{before} oversampling so that copies do not get split across training and test sets.

\begin{figure}
   \centering
    \includegraphics[width=0.9\hsize]{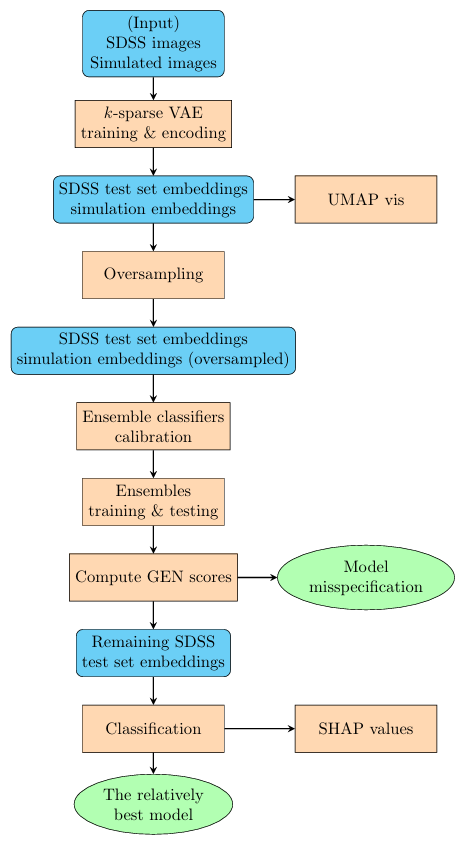}
   \caption{Flow chart of our workflow pipeline. Blue boxes with rounded corners represent data products of the previous step. Orange boxes show the main algorithmic steps in our method. Green ellipses represent output results.}
   \label{fig:pipeline}
\end{figure}
\section{Methods}
\label{sec:methods}

\begin{figure*}
    \centering
    \includegraphics[width=\hsize]{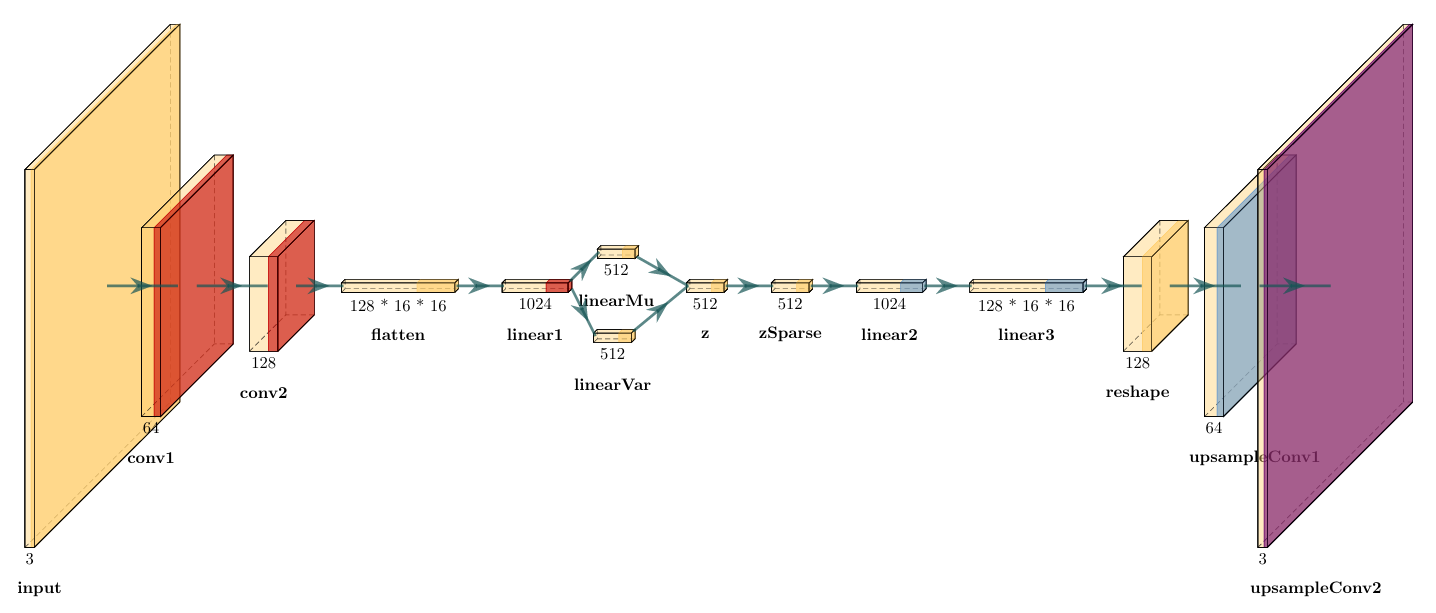}
    \caption{Visualization of $k$-sparse VAE structure. Deep orange stands for LeakyReLU activation function, blue stands for ReLU and purple stands for Sigmoid activation function. The numbers near illustration mean filter sizes.}
    \label{fig:vae-vis}
\end{figure*}

\subsection{Learning summary statistics with limited simulation budgets}
\label{subsec:methods_summary}

In many fields of natural science, models are developed to explain natural phenomena based on some theories. Due to randomness in physical processes, measurement processes and other influence factors, predictions from these models are not deterministic and we obtain instead statistical distributions of parameters. In most cases, we are interested in performing inference on model parameters $\theta$ given observations $\mathcal{D}$, that is, we seek to recover the parameter posterior $p(\theta \mid \mathcal{D})$. Bayes' proportionality $p(\theta \mid \mathcal{D}) \propto p(\mathcal{D} \mid \theta)p(\theta)$ connects the posterior $p(\theta \mid \mathcal{D})$ with the the likelihood $p(\mathcal{D} \mid \theta)$ and the prior $p(\theta)$. 

The data generation process can be described by sampling some unobserved internal states $y$ according to $y \sim p(y \mid \theta)$. The likelihood function is then calculated by marginalizing the joint distribution $p(\mathcal{D}, y \mid \theta)$ over all possible internal states $y$:
\begin{equation}
    p(\mathcal{D} \mid \theta) = \int p(\mathcal{D}, y \mid \theta) dy
\end{equation}
However, even in the case of relatively simple models, this integral is usually intractable. However, through implicit likelihood models \citep{cranmer2020frontier} also known as simulation-based inference (SBI) we can still approach these problems. Given a model, a simulator is created to produce simulated data from model parameters $\theta$ for the phenomena of interest: $samples \sim simulator(\theta)$. Given a sufficiently fast simulator, we can then make use of a lot of data parameter pairs generated by the simulator to learn the connection between the two and hence enable inferring model parameters from data.

Our model comparison problem is a variant of SBI. The difference is that we are interested in the model posterior $p(M|\mathcal{D})$ conditioned on data $\mathcal{D}$ rather than the posterior of model parameters. 

A typical approach in SBI methods for model comparison is reducing the original data into fixed summary statistics (also called embeddings in our context) to avoid working with high-dimensional observables, such as galaxy images. 
Additionally, \citet{radev2021amortized} proposed to train embedding networks that capture the structure of the original data, avoiding catastrophic information loss and biased results caused by hand-crafted summary statistics \citep{robert2011lack, marin2018likelihood}. 
However, end-to-end learning of summary statistics requires large simulation budgets that are infeasible in our setting. For instance, running the TNG100 simulations alone on the Cray XC40 Hazel Hen supercomputer \footnote{\url{https://www.tng-project.org/people/}} demands 1.5 years of runtime on several ten thousand cores, equivalent to millions of CPU hours, making such simulation efforts prohibitively expensive for different model variants. To overcome this problem, we take another approach and leverage the large body of real observational data from SDSS to train an embedding network (i.e., an encoder) in a fully unsupervised manner as part of an information maximizing variational autoencoder architecture.

After training on observational images, we ``freeze'' the encoder and embed the simulated images into the lower dimensional latent space.
This small dataset of ``labeled'' embeddings then serves as the training data for an ensemble of classifiers. Once training has converged, we apply the trained ensemble classifiers to the embeddings of SDSS test set galaxies and perform out-of-distribution detection to find those SDSS images that cannot be accounted for by any of the simulators (i.e., due to model misspecification). Finally, we discard this part of SDSS images in our final model selection task and perform amortized Bayesian model comparison only on the subset of ``in-simulation'' embeddings. Only in this way we obtain trustworthy model posterior probabilities.

In the next subsections we explain in detail each step of our methods. To better illustrate our workflow, we show a flow chart of our pipeline in \autoref{fig:pipeline}.

\subsection{Auto-encoding galaxy images}
\label{subsec:vae}

We use a $k$-sparse variational autoencoder \citep[VAE;][]{kingma2013auto} based on the $k$-sparse autoencoder \citep{makhzani2013k} to encode galaxy images to embeddings. Compared to plain autoencoders, VAEs provide a probabilistic framework by encoding original data to a distribution instead of single points and help prevent overfitting. Additionally our approach of employing sparsity in the VAE helps us achieve better disentanglement in the latent space, thus facilitating physical interpretability in a later step.

In practice, we sample an embedding $z$ from the distribution of embeddings by using the reparameterization trick - a technique that allows gradients to flow through random variables by expressing them as a deterministic function of the distribution's parameters and a noise term. And we incorporate the MMD-VAE loss from the InfoVAE family \cite{zhao2017infovae}, where MMD stands for Maximum Mean Discrepancy, to avoid common problems with the standard VAE KL divergence loss (Kullback-Leibler divergence) and encourage maximally informative compression. Our final loss function is 
\begin{equation}
    \mathcal{L} = \mathbb{MMD}^2(q_\phi(z)\,\|\,p(z)) + \text{MSE}(x_{\rm{recon}}, x)
\end{equation}
where $q_\phi(z)$ represents the approximate distribution of the embedding $z$, $p(z) \sim \mathcal{N}(0, \mathbb{I})$, $\text{MSE}$ stands for the mean squared error, $x$ is the input image and $x_{\rm{recon}}$ is the image reconstructed by the decoder. Here we set a 1:1 ratio between the MMD and MSE terms, following the official implementation of InfoVAE. Additionally, we explored the impact of varying this ratio by testing two alternative configurations: $\mathcal{L} = 0.1 \times \mathbb{MMD}^2(q_\phi(z) \| p(z)) + \text{MSE}(x_{\rm{recon}}, x)$ and $\mathcal{L} = \mathbb{MMD}^2(q_\phi(z) \| p(z)) + 0.5 \times \text{MSE}(x_{\rm{recon}}, x)$, to assess the robustness of our results. However, under both conditions, our findings and conclusions remained consistent, and as such, we do not present these additional results in \autoref{sec:results}.

During training, we compute the embedding $z$ in the feed forward phase, then sparsify it by keeping only the $k$ largest activations (absolute values) and setting the rest to zero. The computation of the loss function and the input of the VAE decoder both use the sparsified $z$. 
We train the $k$-sparse VAE on the SDSS training set with dimension of $z$ equal to 512 and $k = 32$. The embedding dimension and sparsity level $k$ are hyper-parameters that can be tuned. We choose the sparsity ratio to be $32 / 512 = 0.0625$ for our $k$-sparse VAE achieving a balance between capturing local features and global features \citep{makhzani2013k}. Then we encode the SDSS test set and the simulated images to 512 dimensional embeddings with $k = 64$. The sparsity level $k = 64$ is determined by following the result of the original paper, as a larger $k$ during the encoding phase reduces the error rate of downstream classification task \citep{makhzani2013k}. This suggests that different $k$ values for training and testing do not introduce any systematic effects that could resemble OOD behavior; otherwise, the classifier's performance would worsen as it wouldn't handle OOD data well. 
In our setup, the choice of larger $k$ during testing yields more robust embeddings without compromising the disentanglement or interpretability of the latent space. We also verified that our main results are qualitatively stable across a reasonable range of $k$ values (from 16 to 64), indicating that our framework is not overly sensitive to this hyperparameter.

The network structure of our $k$-sparse VAE is illustrated in \autoref{fig:vae-vis}. The encoder consists of convolutional layers, flatten layers and linear layers. In the decoder, we employ an upsampling layer followed by a convolutional layer, rather than using a deconvolution layer, in order to mitigate the checkerboard effect. The checkerboard effect refers to a visual artifact in reconstructed images, characterized by abrupt, unnatural transitions in pixel colors or luminance values, which result in a grid-like pattern resembling a checkerboard.

\subsection{\texorpdfstring{Training and encoding of \( k \)-sparse VAE}{Training and encoding of k-sparse VAE}}
\label{subsec:vae_results}

\begin{figure}
    \centering
    \includegraphics[width=\hsize]{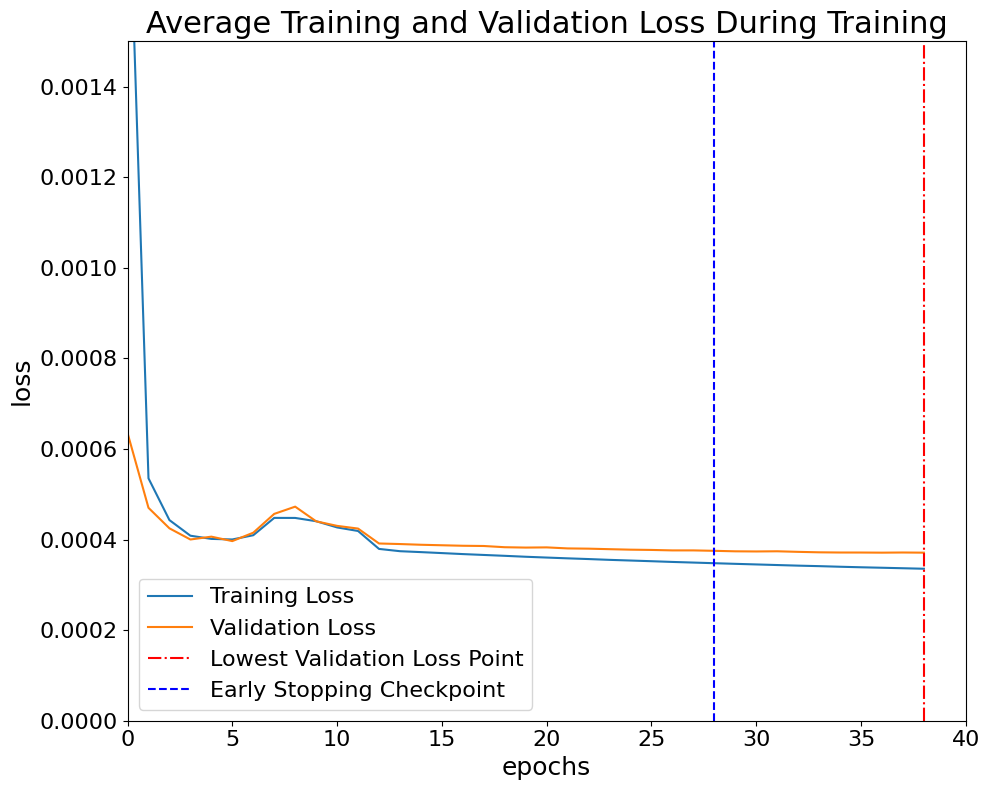}
    \caption{Average training and validation loss of each epoch during $k$-sparse VAE training on the SDSS training set and validating on the SDSS early stopping validation set. The vertical red line shows the lowest validation loss checkpoint and the vertical blue line indicates the epoch where our VAE model is saved. This early stopping mechanism avoids overfitting.}
    \label{fig:vae-avg-loss}
\end{figure}

As suggested by \citet{makhzani2013k}, we deploy a scheduling of the sparsity level during training by starting with a larger sparsity level $k = 81$ and then linearly decrease it to the target sparsity level $k = 32$ in the first 7 epochs. By doing so, we avoid the problem of ``dead hidden units'' that may appear in the $k$-sparse autoencoder, where some hidden units are selected during the initial epochs and reinforced in later ones, while others remain unadjusted. For training the $k$-sparse VAE on the SDSS training set, we use the Adam optimizer \citep{kingma2014adam} with an initial learning rate of $10^{-3}$ and a reduce-on-plateau schedule for dynamically reducing the learning rate by a factor of $0.1$ if the average validation loss per epoch has no improvement after 5 epochs. We have an early stopping mechanism in place, where the training halts if the average validation loss does not decrease for 10 consecutive epochs in which case we choose the model checkpoint at those previous 10 epochs. Using a batch size of 400, the training of the $k$-sparse VAE for 38 epochs on a single A100 GPU takes 6,802 seconds ($\sim1$ hour $53$ minutes). The final model checkpoint which we evaluate is hence at the 28th epoch. The average training and validation loss of each epoch during training is shown in \autoref{fig:vae-avg-loss}.

\subsection{Dealing with imbalanced dataset}
\label{subsec:oversampling}

As stated in \autoref{subsec:simulated_images}, our simulation data from 6 different simulation models form an imbalanced dataset where TNG100 is the majority class and all other models are minority classes. Imbalanced datasets are a common and challenging problem which often results in a biased machine learning model that performs well on the majority classes but poorly on the minority classes. 
To alleviate this issue, we oversample the images in the minority classes by copying them once. In our experiments, without oversampling, the classifiers in later steps are poorly calibrated with suboptimal confusion matrices and calibration curves.

We are aware that the usual procedure in oversampling is to create a balanced data set by replicating the minority instances according to the class ratios. In our experiments, if images of minority classes are copied more than once, then the calibration curve becomes worse (see the classifier calibration section later).
We have also considered modifying these copies (e.g. geometric transformations, color space transformations, etc). However, there are physical meanings behind color and shape of galaxies, so we cannot change these properties randomly.

While undersampling the TNG100 class could in principle balance the dataset, we chose not to do so in order to preserve the full diversity of available simulated galaxy images, which is especially important given the limited sample sizes across all models. Reducing the majority class would have discarded valuable information and likely weakened classifier performance and calibration. Additionally, although standard augmentation techniques such as rotations and flips are common in image classification, we deliberately avoided them in our setup. The simulated images include structured observational effects (see \autoref{subsec:simulated_images})—such as directional PSF convolution and realistic sky noise—which are sensitive to spatial orientation. Applying such augmentations would introduce unphysical artifacts that are not present in SDSS observations and could lead to spurious classification features or affect OOD detection. We therefore adopt a conservative oversampling strategy by duplicating each minority class once, which we found to improve classifier calibration while maintaining physical fidelity in the training set.
 
In addition, oversampling of embeddings does not work in our case. Each embedding corresponds to a simulated galaxy image. Using resampling techniques like SMOTE \citep[Synthetic Minority Over-sampling Technique][]{chawla2002smote} may create embeddings that match unrealistic images.

\begin{figure*}
    \centering
    \includegraphics[width=\textwidth]{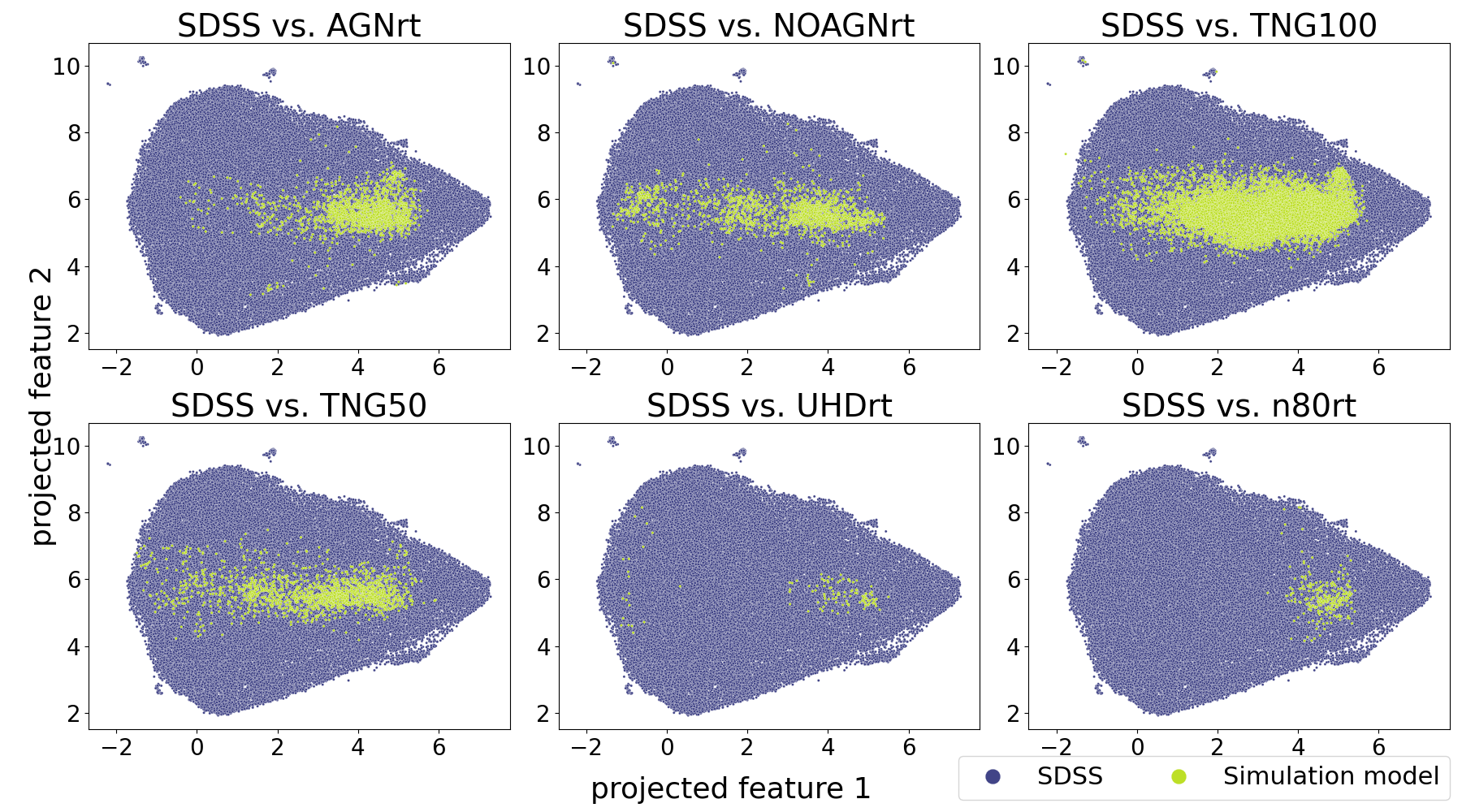}
    \caption{UMAP projection of simulated latent embeddings (yellow) compared to the SDSS test set (purple).}
    \label{fig:umap}
\end{figure*}

\subsection{Amortized Bayesian model comparison and model misspecification detection}
\label{subsec:bmc}

Bayesian model comparison (BMC) offers a probabilistic / relative comparison of multiple models rather than an absolute assessment of any individual model. BMC compares candidate models by evaluating the ratio of their posterior probabilities given known data and prior knowledge. It can be cast into a classification task by training a classifier with a proper loss \citep{gneiting2007strictly} to induce a categorical distribution over the model indices $\mathcal{M}$ given the observed data $D$ \citep{radev2021amortized}:
\begin{equation}
    \mathcal{M} \sim p(\mathcal{M} \mid \mathcal{D}) \propto p(\mathcal{D} \mid \mathcal{M})\,p(\mathcal{M})
\end{equation}
Correspondingly, we train an ensemble of classifiers on the ``labeled'' embeddings of the simulated images and then use the trained ensemble to estimate posterior model probabilities from the SDSS test set. The model with the best relative fit from a Bayesian perspective \citep{mackay2003information, radev2021amortized} is then the one that is preferred by the classifier. 
The classifier can be efficiently reused for inference as new observations come in, hence the training cost \textit{amortizes} over multiple observations. This is a notable advantage compared to posterior predictive methods like WAIC \citep[widely applicable information criterion][]{watanabe2013waic} and cross-validation based methods \cite{vehtari2017practical}, which need to calculate the likelihood of each observed data point given each model is required.

However, we cannot simply apply the ensemble to all embeddings of our SDSS test set since some of them may be out-of-distribution \citep[OOD;][]{yang2024generalized} relative to the simulations, which can lead to incorrect or unstable predictions \citep{schmitt2023detecting, elsemuller2023sensitivity}.
Intuitively speaking, when classifying a data point that lies far from the distributions of all candidate classes, classifiers may assign a disproportionately high probability to the class that is only slightly closer to the data point.

In our context, OOD embeddings occur when the simulations differ significantly from the (actually observed) SDSS test data and thus indicates model misspecification, which shows that the simulation differs significantly from the true data generating process (also called simulation gap).
We can use any \textit{post hoc} OOD score \citep{yang2024generalized} to \textit{detect} observations for which the models are misspecified. A \textit{post hoc} OOD score identifies OOD samples using a trained model's outputs without retraining or altering the model. In practice, we perform out-of-distribution detection using the Generalized ENtropy score \citep[GEN score][]{liu2023gen} which is defined as: 
\begin{equation}
    \smash{G_{\gamma}(p) = \sum_j p_j^{\gamma} (1-p_j)^\gamma}
\end{equation}
with $\gamma \in (0, 1)$, where $p$ are the probabilities of all classes calculated by applying the softmax function to the logits produced by the classifiers. In practice, we follow the recommendation from \citep{liu2023gen} in setting $\gamma = 0.1$, which was shown to perform well across diverse classification tasks. We additionally verified that our downstream results—specifically, the classifier calibration and relative model rankings—remain qualitatively stable for alternative values of $\gamma$. This indicates that our OOD detection method is not overly sensitive to the specific choice of $\gamma$.

We consider all simulation models in the computation of the GEN score since we only have 6 classes rather than hundreds of classes as in the paper of \citep{liu2023gen}. Finally, following the implementation of the original paper \citep{liu2023gen}, we compute negative GEN scores.
In order to perform OOD detection, we proceed as follows: We fit a classifier to the SDSS test set and compute the corresponding GEN score distribution. Similarly, we compute the reference GEN score distribution by classifying the simulation test set. If the GEN score distribution of the SDSS test set lies significantly outside of the reference GEN score distribution, then the SDSS test set is OOD, implying that the simulations deviate from reality. In this case, we perform model comparison only on the subset of SDSS data that ``agree'' with the simulation embeddings.
For this, we take the percent point corresponding to 95\% of the reference GEN score distribution as a threshold and ignore all SDSS latent embeddings with a GEN score beyond this threshold. To this ``cleaned'' SDSS test dataset we apply the classifiers once again to derive our final model posteriors. In this way, we can increase the robustness of model posterior estimates and the corresponding theoretical implications. Additionally, in \autoref{subsec:ood_threshold_analysis} we conduct a sensitivity analysis on the threshold of the reference GEN score distribution to justify our choice of 95\% as a reasonable value.

We like to note here that by using the GEN score to select SDSS galaxies that are in agreement with the simulation models circumvents the need for any simplified cut in SFR or total stellar mass. Using our embedding vectors combined with a GEN score actually allows for a more nuanced comparison between simulated and real galaxy images than any simple cut in stellar mass or SFR would have accounted for. Additionally, since we pre-train our embedding network on real images, any type of distribution shift not only a shift in stellar mass or SFR but also a shift in a miss-match in noise patterns or PSF of the simulated images will lead to the simulation data being out-of-distribution compared to real galaxy images. This is the key motivation for doing model comparison only on in-distribution data. Combined with our explainable AI approach described in section \ref{subsec:shap} our method is able to investigate the reasons for out-of-distribution data and should reveal miss-matches in PSF or noise patterns.

\begin{figure*}
    \centering
    \subfigure[random forest]{
        \includegraphics[width=.32\textwidth, trim={.1cm 0cm 1.6cm .65cm}, clip]{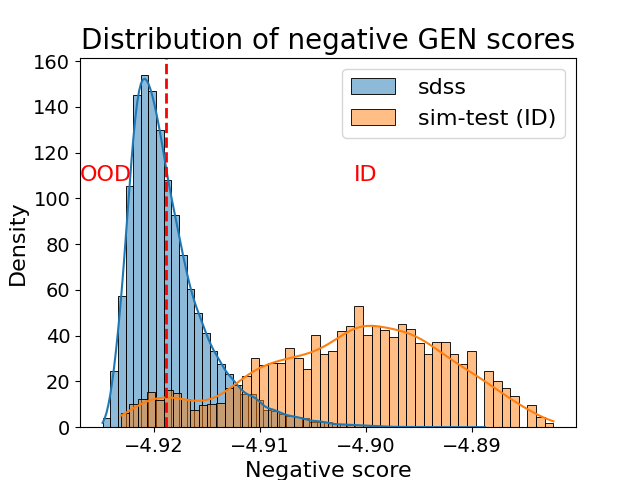}
    }
    \subfigure[XGBoost]{
        \includegraphics[width=.32\textwidth, trim={.1cm 0cm 1.6cm .65cm}, clip]{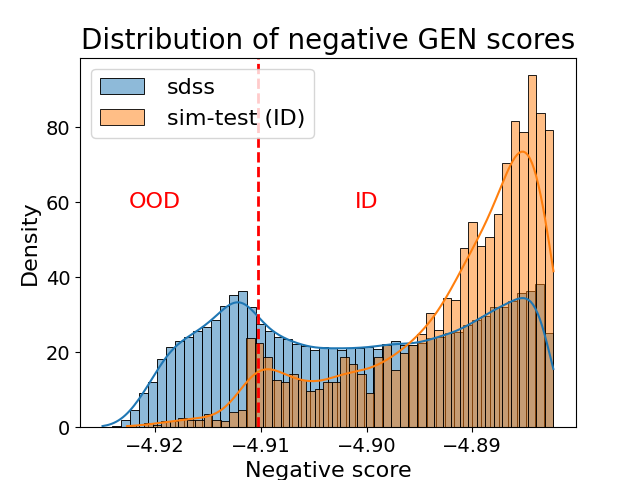}
    }
    \subfigure[stacking-MLP-RF-XGB]{
        \includegraphics[width=.32\textwidth, trim={.1cm 0cm 1.6cm .65cm}, clip]{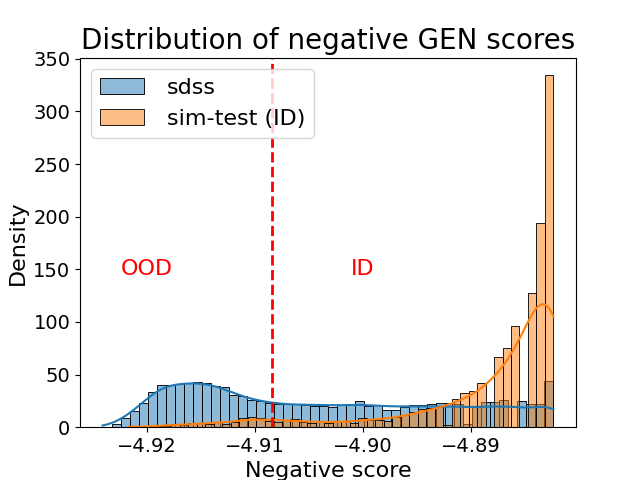}
    }
    \caption{GEN score distribution of all simulation models test set (orange) and SDSS test set (blue). Left: random forest, Middle: XGBoost, Right: stacking-MLP-RF-XGB. The red vertical line represents the 95\% threshold, with the region to its right corresponding to in-distribution data and the region to its left indicating out-of-distribution (OOD) data.}
    \label{fig:gen-ood}
\end{figure*}

\subsection{Ensemble classifiers and calibration}
\label{subsec:ensemble}

As a baseline model we choose a random forest classifier \citep{breiman2001random}, which has been used in previous model comparison papers \citep{pudlo2016reliable, marin2018likelihood}. As stated before, we have a small set of simulation data and ensemble methods handle this situation better than a single neural network. We also train an XGBoost \citep[eXtreme Gradient Boosting][]{chen2016xgboost} classifier and we additionally train a stacking ensemble classifier (referred to as stacking-MLP-RF-XGB in the following), which combines 3 base classifiers (multilayer perceptron, random forest and XGBoost) with a random forest serving as the final meta estimator. 

In the random forest classifier, we set the number of trees to 100 and class weight to balanced subsample, which means the weight assigned to each class is determined by its inverse proportionality to the frequency of the class within the bootstrap sample used for growing each tree. We use the XGBoost classifier from the XGBoost library \citep{chen2016xgboost}, in which we use the multiclass softmax objective and hist tree method, which is the fastest approximated training algorithm. For stacking-MLP-RF-XGB classifier, the parameters of random forest and XGBoost are the same as described above. For the MLP, we set two hidden layers: the first one with 128 neurons and the second one with 64 neurons. And we set the activation function to ReLU, use the Adam optimizer \citep{kingma2014adam} with an L2 regularization alpha set to 0.01, and maximum iteration of 300. 

The training of our stacking classifier proceeds as follows: Each of the 3 base classifiers are trained to output a 6-dimensional class probability vector for each instance. Then we concatenate the class probability predictions from each of the 3 base estimators to obtain an 18-dimensional vector as the meta-feature vector (input feature for meta classifier) for each instance. Finally, the meta classifier (random forest) is trained on this new dataset with 18 features.

We perform the classifier calibration by doing twice repeated stratified $5$-fold cross-validation on the simulation training set. In practice, we use the calibration curve function from the bayesflow library\footnote{\url{https://bayesflow.org}} \citep{radev2023bayesflow}. It integrates the computation of expected calibration error (ECE) of a model comparison network proposed by \citet{naeini2015obtaining}. We can better explain the calibration curve by describing its creation, which involves the following steps for a simulation model:
\begin{itemize}
    \item[\textendash] The probability range [0,1] is divided into 10 bins of equal length. Then we assign predicted probabilities to bins.
    \item[\textendash] In each bin, we compute the mean of predicted probabilities and this is the value on the x-axis.
    \item[\textendash] In each bin, we compute the proportion of data that actually belongs to this model and this is the value on the y-axis.
\end{itemize}

From the calibration curves in \autoref{fig:cc} in \autoref{sec:appendix_calibration}  we can see that the stacking MLP-RF-XGB classifier is the best one with lowest ECE score and calibration curves close to the optimal diagonal line. Overall, the stacking-MLP-RF-XGB classifier achieves better recovery than random forest or XGBoost. The second best calibrated classifier is XGBoost which performs better than random forest. Note that the calibration curves of all classifiers for UHD and n80 are not well calibrated. Since these two classes are the smallest, we attribute this simply to the lack of data.
This limitation highlights a general requirement of our approach: while the method is designed to handle relatively small simulation datasets, reliable calibration typically requires a minimum of several hundred images per model class to ensure meaningful posterior estimation. For extremely small datasets classifier confidence and interpretability can degrade.

The resulting figures for the confusion matrices are shown in \autoref{fig:cm} also in \autoref{sec:appendix_calibration}. Each of the 3 classifiers produces a satisfying confusion matrix. All classifiers achieve a very high accuracy of classifying AGN, NOAGN and TNG100, while the accuracy of classifying TNG50, UHD and n80 is lower. Again, for UHD and n80, this may be due to the lack of data. Notice that the accuracy of classifying TNG50 is relatively poor in the sense that classifiers tend to falsely predict some data of TNG50 as TNG100. However, this is actually a reasonable result since the only difference between TNG50 and TNG100 is the higher resolution of TNG50. Hence we hypothesise that this makes the two less distinguishable.

Considering both confusion matrix and calibration curve, we conclude that we should trust the stacking-MLP-RF-XGB classifier most in the final classification results. 

Finally, the random forest is trained on CPU (2 x 32-Core AMD Epyc 7452) while the XGBoost classifier is trained on a GPU (A100). Among the base estimators of stacking-MLP-RF-XGB, XGBoost is trained on GPU while others are trained on CPU. The training of random forest takes 12.6 seconds, XGBoost takes 14.6 seconds and stacking-MLP-RF-XGB takes 293.8 seconds.

\subsection{Physical insights through explainable AI}
\label{subsec:shap}

SHAP \citep[SHapley Additive exPlanations]{lundberg2017unified} is an XAI (explainable machine learning) method used to explain the contribution of each feature to the final prediction of a machine learning model. It borrows the concept of Shapley values from game theory which are designed to distribute total gain among players in a group based on each one's contribution. For a game, the Shapley value for player $i$ is defined as
\begin{equation}
\label{eq:shapley}
    \phi_i = \sum_{S \subseteq N \setminus \{i\}} \frac{|S|! \, (|N| - |S| - 1)!}{|N|!} \left( v(S \cup \{i\}) - v(S) \right)
\end{equation}
where $N$ is the set of all players, $S$ is a subset of players not including player $i$, $|S|$ is the number of players in subset $S$, and $v(S)$ represents the total payoff of coalition $S$ resulting from the cooperation of players. Intuitively, the Shapley value computes the average marginal contribution of a player to all possible coalitions.

In XAI the SHAP method adapts \autoref{eq:shapley} by converting players to features and the value function $v(S)$ now becomes the model prediction function. In this sense, the SHAP value of feature $i$ is the average marginal contribution of the feature to the prediction, accounting for all possible combinations of other features. Obviously, the exact computation of \autoref{eq:shapley} is not feasible and hence in practice SHAP values are estimated \citep{lundberg2017unified}. For an instance vector $x$, the sum of SHAP values of all features is equal to the difference between the actual prediction and the average prediction:
\begin{equation}
    \sum_{j=1}^n \phi_j = f(x) - \mathbb{E}_x(f(x))
\end{equation}

In order to compute the expected value of the model prediction the SHAP software uses ``background data'' to calculate a reference point for the SHAP value interpretation. Thus, the background data should be chosen to represent the data distribution, which is typically selected from the training dataset. Therefore, we used stratified sampling to select 50\% of the simulation training set as background data due to the imbalanced datasets. Then we apply the trained SHAP tree explainer to 60\% randomly sampled ``in-distribution'' SDSS test data for balancing calculation time and result reliability. 

In \autoref{subsec:shap_results} we interpret our model comparison results qualitatively in light of physical difference between the simulation models via an analysis of SHAP values on the XGBoost classifier. In practice, we use a variant of the original SHAP method designed for tree-based machine learning models called TreeSHAP \citep{lundberg2020local}. We like to note that the SHAP method does not work for complex stacked classifiers such as our preferred stacking-MLP-RF-XGB. However, all classifier results presented and tested here are qualitatively the same and our insights gained from XGBoost will carry over to the more complex classifiers.

\section{Results}
\label{sec:results}

\subsection{Latent space visualization using UMAP}
\label{subsec:umap_vis}

We visualize the latent space embedding using UMAP \citep[Uniform Manifold Approximation and Projection][]{mcinnes2018umap} to project the 512-dimensional embeddings into a 2-d space in \autoref{fig:umap}. We train the UMAP model solely on the SDSS test set to obtain the corresponding embeddings and apply it then to each simulation to visualize the relative positions of simulated data and SDSS test set. By doing so we can get an intuition of the gap between the simulation models and reality.

From \autoref{fig:umap} it is clear that the embeddings from the 6 simulation models overlap only with a small part of the SDSS test data which implies that all simulation models can only explain a small fraction of observed galaxies. The difference in simulation models that appear are mainly due to differences in galaxy populations and not in physical resolution of the simulation models. For example, IlustrisTNG50 models a smaller volume compared to IlustrisTNG100 and hence misses a couple massive galaxies that are included in IlustrisTNG100. The same is true for NIHAO-UHD vs. NIHAO-noAGN.

This is further confirmed by the GEN score distributions which are generally of different shape for the observational dataset and the simulation models (see \autoref{subsec:ood_results}). This result suggests that all simulation models are somewhat misspecified and our approach opens up various avenues to improve upon the simulation models using explainable AI methods, such as SHAP values (see our results in \autoref{subsec:shap_results}).

\subsection{Model misspecification detection via OOD}
\label{subsec:ood_results}

As explained in \autoref{subsec:bmc} we utilize GEN scores for OOD detection.
The resulting figures for the GEN score distributions are shown in \autoref{fig:gen-ood}. As we described before, we compute GEN score distribution of SDSS test set (blue distribution) and simulation model test set (orange distribution). We define the threshold as the percent point corresponding to 95\% (for a sensitivity analysis on this threshold see \autoref{subsec:ood_threshold_analysis}) of the reference GEN score distribution. More specifically, SDSS embeddings with GEN score left to this threshold on the \autoref{fig:gen-ood} are detected as OOD data and right to this threshold are defined as in-distribution data. Looking at the fraction of OOD data for each classifier, we have 55\% in-distribution and 45\% OOD SDSS data for the stacking-MLP-RF-XGB classifier, 42\% in-distribution and 58\% OOD SDSS data for the random forest and 71\% in-distribution and 29\% OOD SDSS data for XGBoost. 

We note that the GEN score distributions differ across classifiers, which can be attributed to differences in their probability calibration behavior. Random forests often produce overconfident class assignments due to their discrete tree-based structure, while XGBoost tends to yield more flexible but sometimes still overconfident predictions. The stacked ensemble classifier combines multiple base estimators, leading to better-calibrated probabilities and smoother GEN score distributions. Since the GEN score is computed from the class probability outputs, such calibration differences naturally propagate into differences in OOD detection outcomes. Importantly, despite these differences, the relative model ranking and classification outcomes remain qualitatively stable across classifiers and
all 3 ensemble classifiers show a consistently high OOD level, indicating that all simulation models are somewhat misspecified.
Since we focus here on model comparison an in depth analysis of the detailed origin of this model misspecification is beyond the scope of this paper and will be investigated in future work. However, it worth noting that this high OOD fraction can have several reasons beyond just physical model inaccuracies. A mismatch in redshift range, size, SFR or stellar mass can also lead to a high OOD fraction and we deliberately decided to not apply hand-crafted cuts on these parameters. 
Instead, we apply non-parametric cuts via the GEN score to allow for a more nuanced selection of galaxy images for the model comparison task.

We note, however, that a high OOD fraction does not necessarily imply that the simulation models are physically inaccurate. It may also arise due to limited coverage of the observational parameter space by the simulations. For instance, our simulated galaxy images were all generated at a fixed redshift of 0.109, matching the mean redshift of the SDSS sample. As a result, the simulations lack galaxies with smaller apparent sizes that would occur at higher redshifts. Additionally, the selection functions in stellar mass and SFR differ between the SDSS and the simulations. While we deliberately avoid imposing matching cuts in these parameters—favoring instead a latent-space-based, non-parametric OOD filtering—we acknowledge that this design choice contributes to the observed simulation gap. A promising direction for future work would be to forward-model simulated galaxies across the full redshift range of SDSS by varying their apparent size and observational conditions accordingly. Such an approach may enhance the overlap in latent space and reduce the OOD fraction.

The in-distribution SDSS subset selected via GEN score thresholding is not based on morphology or color alone, but rather reflects latent features learned from the full image—including structural, photometric, and observational characteristics. The filtered sample thus spans a range of galaxy types that are well-represented in the simulations, rather than exclusively smooth or quiescent galaxies. This avoids unfairly favoring certain models and preserves diversity within the evaluation set.

\begin{figure}
    \centering
    \subfigure[random forest]{
        \includegraphics[width=\columnwidth, trim={1.5cm 0cm 2.5cm 1.cm}, clip]{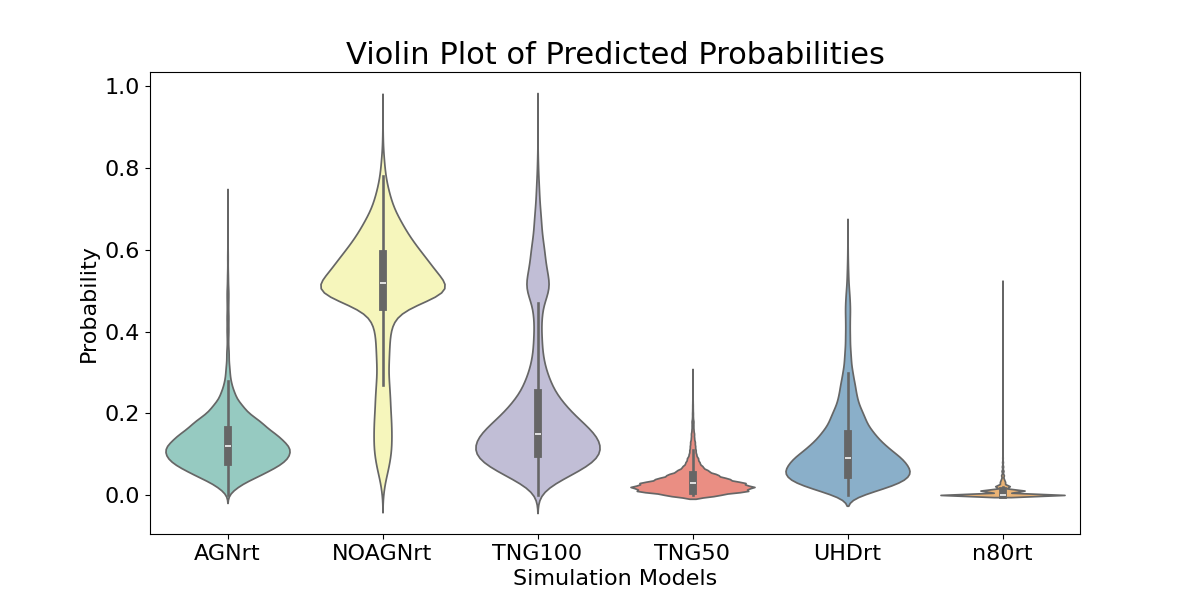}
    }
    \subfigure[XGBoost]{
        \includegraphics[width=\columnwidth, trim={1.5cm 0cm 2.5cm 1.cm}, clip]{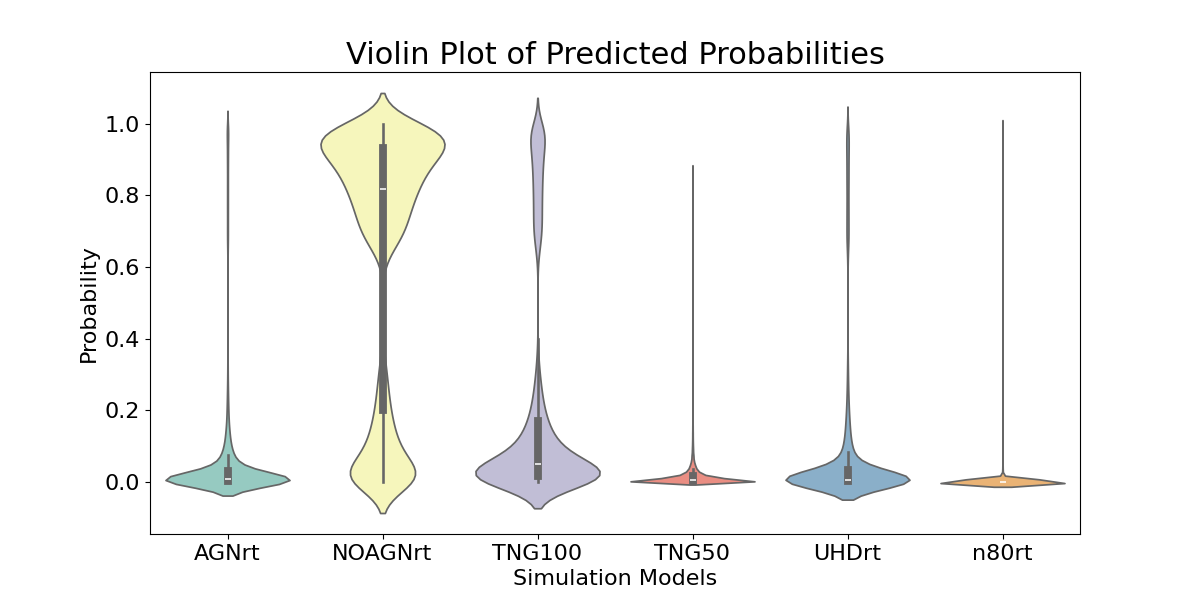}
    }
    \subfigure[stacking-MLP-RF-XGB]{
        \includegraphics[width=\columnwidth, trim={1.5cm 0cm 2.5cm 1.cm}, clip]{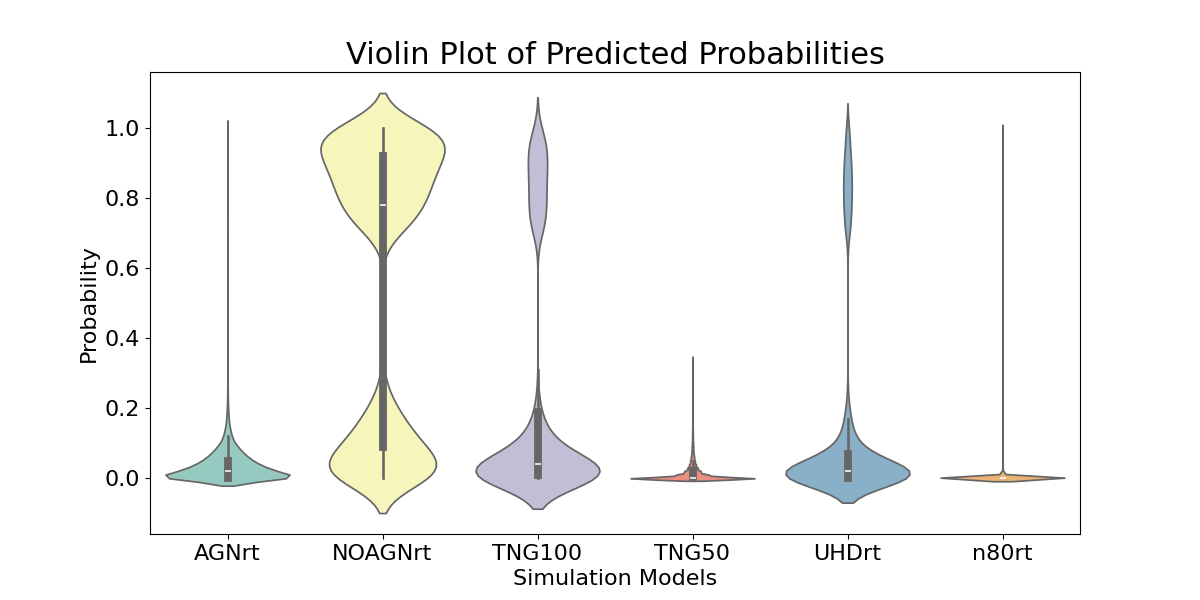}
    }
    \caption{Violin plots show the classification of the in-distribution data of SDSS test set for 3 classifiers. Top: random forest, Middle: XGBoost, Bottom: stacking-MLP-RF-XGB. The white horizontal line inside the box marks the median. The whole gray box shows the interquartile range (IQR), which is defined as
 the difference between the third quartile (Q3) and the first quartile (Q1) of a group
 of data. Hence it shows the middle 50\% of the distribution. And the whisker is calculated as 1.5 $\times$ IQR. It displays the distribution of predicted model probabilities $p(\mathcal{M} \mid \mathcal{D}_i)$ over all data $\mathcal{D}_i$ (y-axis) for 6 simulation models (x-axis).}
    \label{fig:violin}
\end{figure}

\begin{figure*}
    \centering
    \subfigure[NOAGN]{
        \includegraphics[width=0.75\hsize,trim={0cm 2.75cm 3.5cm 2.25cm}, clip]{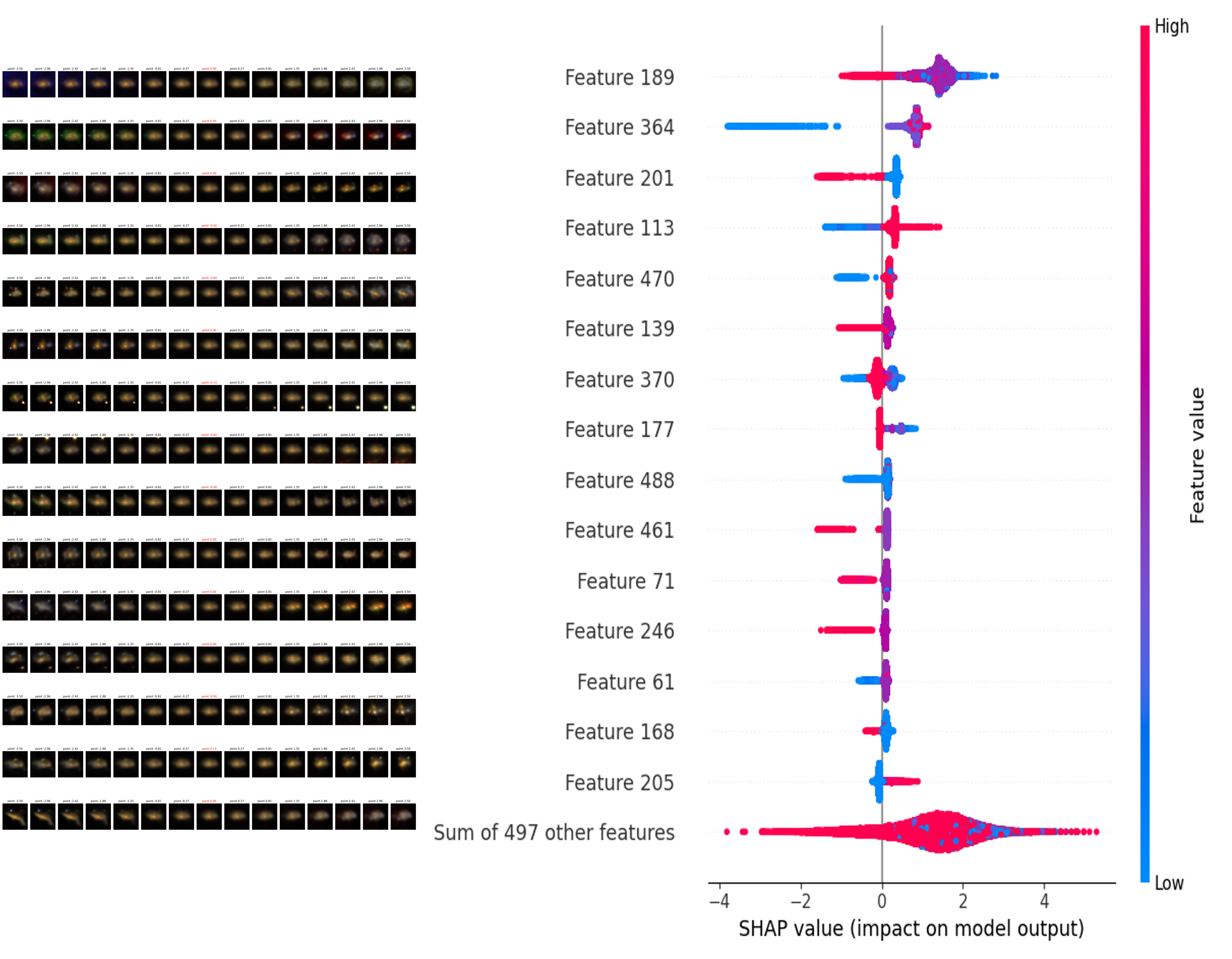}
    }
    \subfigure[TNG100]{
        \includegraphics[width=0.75\hsize,trim={0cm 2.75cm 3.5cm 2.25cm}, clip]{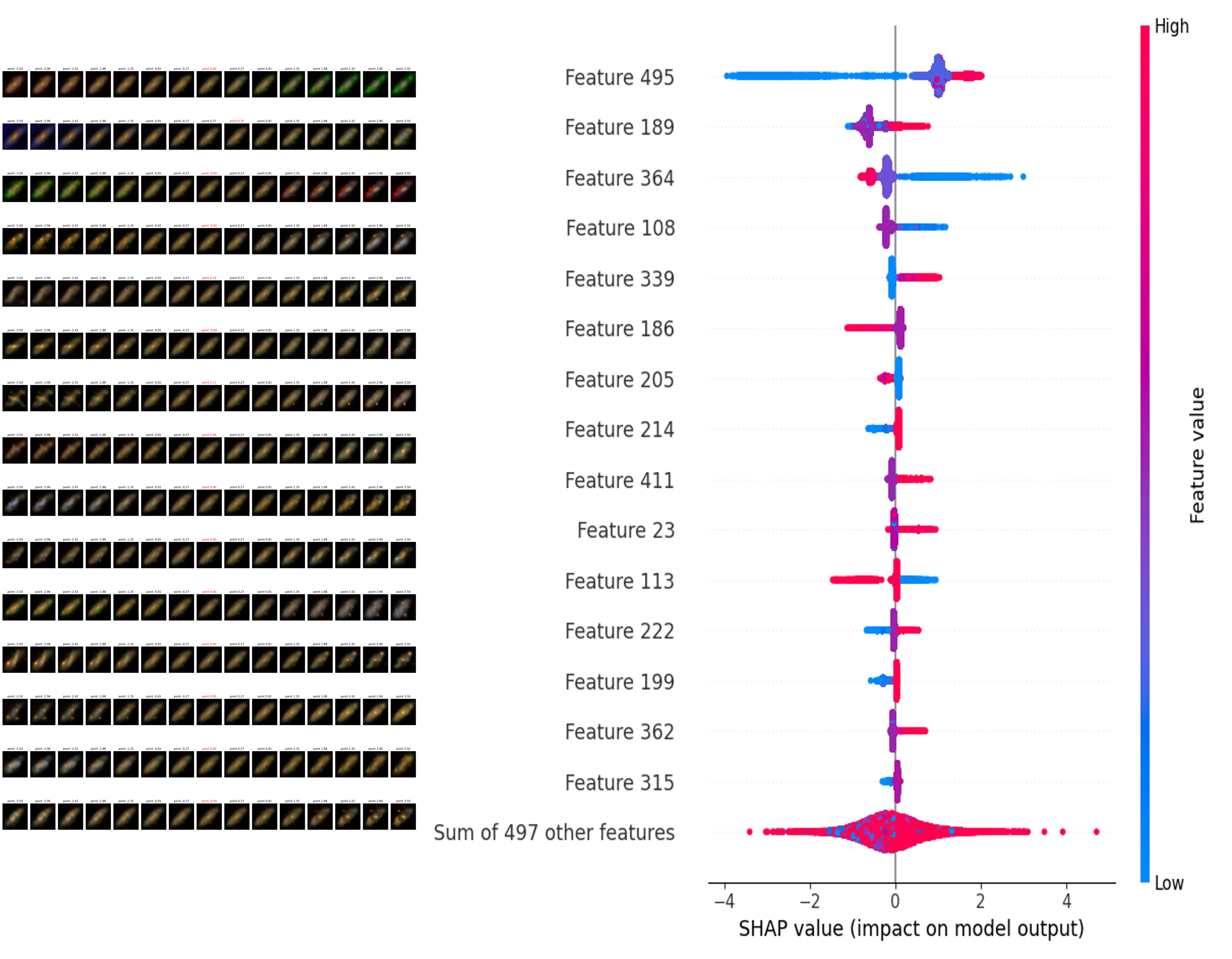}
    }
    \caption{SHAP plots for the XGBoost classifier. Top: NOAGN, Bottom: TNG100. Left: Each row shows a visualization of generated galaxy images varying a single feature value in each column. The middle column indicated by a red title shows the reconstruction from an encoded latent embedding of an example galaxy. To the left (right) we reduce (increase) the value of that entry in the latent embedding varying it by $\sim3\sigma$ around the mean in each dimension.
    Right: $n$-th feature dimension of the latent embeddings ordered by their importance on the prediction of the XGBoost classifier result. The x-axis shows the SHAP value, i.e. the impact of that feature on the prediction while the color coding shows the feature value, e.g. high (red) vs. low (blue) feature values. Each point in the distribution refers to an individual prediction from the test set.
    For example, features 189 (top) and 495 (bottom) show strong color variation while features 470 (top) and 205 (bottom) show strong structural variation.}
    \label{fig:shap}
\end{figure*}

\subsection{Amortized Bayesian model comparison}
\label{subsec:bmc_results}

Applying our Bayesian model comparison pipeline to the in-distribution dataset of the SDSS embeddings, we derive our final result shown in \autoref{fig:violin}. For all classifiers explored here we find rather low model probabilities for AGN, TNG50, UHD and n80. For no AGN and TNG100 we find a bimodal distribution with varying strength depending on the classifier.
The bimodal shape results from the fact that the discarded OOD SDSS test data are primarily those classified between 40\% and 60\% as either NOAGN or TNG100, indicating that it is difficult for the classifier to make a definite prediction for them (see also \autoref{subsec:ood_threshold_analysis}). 

There is a clear preference for the NOAGN model by all 3 classifiers considering the violin shape and the position of box-and-whisker plot inside each ``violin''. This relative preference does not necessarily mean that NOAGN fits the SDSS test set best in an absolute sense. It simply points to the fact that, among all misspecified models, NOAGN generates the most realistic images when compared to all other models. Hence, the model comparison we perform here is a relative comparison across all participating models. 

Note though that also a tiny fraction of the TNG100 and UHD galaxies are well in agreement with SDSS. Interestingly, comparing TNG100/TNG50 and NOAGN/UHD, we find that higher physical resolution does not necessarily provide better agreement with observations.
This might reveal a mismatch between simulation resolution and the employed sub-grid physics which might fail to result in realistic simulations if not adapted for higher resolution.

\subsection{Physical insights via SHAP analysis}
\label{subsec:shap_results}

As mentioned in \autoref{subsec:shap}, we use TreeSHAP from the SHAP software to explain predictions from the XGBoost classifier to gain physical insights into our model comparison results. This should further give us an intuition about the influence of different features on the classification results also for other classifiers explored in this work.

Two examples of the resulting SHAP plots are shown in \autoref{fig:shap} additional plots for all other models are shown in \autoref{sec:appendix_shap_plots}. The overall figure structure is as follows: In the left hand side each row shows a visualization of generated galaxy images varying a single feature value in each column. The middle column indicated by a red title shows the reconstruction from an encoded latent embedding of an example galaxy. To the left (right) we reduce (increase) the value of that entry in the latent embedding. We vary latent embedding entries by about $3\sigma$ around the mean in each dimension.
Note, we keep every other feature value fixed per row and only vary a single feature value. Note, the variation in each dimension encodes the behavior of all encoded galaxy vectors and we display qualitatively what happens to a representative galaxy (a mean galaxy) if its encoding vector is varied within 3$\sigma$ of the data distribution. Hence, the images on the left aid the interpretation of the SHAP feature vectors and are representative of the behavior on the entire dataset not just one example galaxy shown there.

The right part of \autoref{fig:shap} shows the result of the SHAP analysis. Feature $n$ in the right panel corresponds to the $n$-th dimension of the latent embeddings ordered by their importance on the prediction of the XGBoost classifier result. The last row in the right part refers to the impact of all other features combined. The x-axis shows the SHAP value, i.e. the impact of that feature on the prediction while the color coding shows the feature value, e.g. high (red) vs. low (blue) feature values. Each point in the distribution refers to an individual prediction from the test set.

Looking at panel a of \autoref{fig:shap}, we find that feature 189 has the strongest impact on the prediction. Looking closer at the distribution of SHAP values and its color distribution, we see that feature 189 generally has a positive impact on the prediction. More nuanced, we find that if its feature value is lower (bluer) the impact is stronger and less strong in case of large feature values. It can even have a negative impact on classification in case of very large feature values (red). This result can now be interpreted by investigating the left side of this plot to gain insights into what feature 189 physically does. Looking at the visualizations of images generated by varying feature 189 we find that a low (high) feature value corressonds to more red (more blue) galaxies while at the same time the central light concentration decreases with increasing feature value. 

The physical interpretation of this finding is hence as follows: SDSS galaxies are more likely to be classified as NIHAO noAGN if their color is redder and the central light concentration is stronger. This can be interpreted that in general NIHAO noAGN galaxies are redder and more concentrated. This is completely contrary to the findings for TNG100 shown in panel b of \autoref{fig:shap}.

Overall, since the feature dimensions are ordered by their importance on the prediction of the XGBoost classifier result, the upper and lower panels in \autoref{fig:shap} have slightly different rankings and show different feature vectors. However, feature 364 and 189 are similarly important for the NOAGN and TNG100 classification. Additionally the importance of feature 113 and 205 are also shared between the two models.
Interestingly, the effect of feature 364 and 189 on the classification output is exactly opposite - for NOAGN (TNG100) these features have an overall positive (negative) impact on the classification if they show a large feature value. Looking at the meaning of these two features, we find that 189 strongly correlates with color where a low feature value represents more red galaxies and a larger value encodes bluer ones. Similarly, feature 364 encode green to red galaxies where additionally the substructure inside the galaxy varies with feature value, the larger (smaller) this feature the more red (green) spots appear in the galaxies. From this we conclude that NOAGN tend to be redder and clumpier than TNG100 which, in turn, are bluer and smoother by comparison.
This difference might point towards different star formation histories and present day star formation rates since younger stars are on average bluer. A similar conclusion can be drawn from the other two common features 205 and 113.
Hence, we conclude that BMC combined with SHAP value analysis is suited to interpret model comparison results and to gain insights into the physical meaning of the results. This offers opportunities to not only explain model results but further help to improve the physical realism of galaxy formation simulations by pointing out physical features that match or don't match specific observational data.

We like to note here that physical interpretation through SHAP values is based on interpretable and disentangled feature dimensions in the embedding space. Through our choice of a $k$-sparse VAE we have used a sufficiently capable encoder algorithm to disentangle the embedding features. However, we do not expect perfect disentanglement as already from a physical perspective we would expect some kind of feature correlation as we e.g. already know that elliptical galaxies are redder and older compared to spiral galaxies that are bluer and younger.


\section{Discussion}
\label{sec:disucssion}

While our framework provides a robust, explainable, and computationally efficient pipeline for simulation-based model comparison, several limitations and caveats merit discussion:

\paragraph{Numerical effects and resolution sensitivity:}

Our approach is inherently image-based and thus may be sensitive to both image and simulation resolution. All images here are resampled to a fixed resolution of $64 \times 64$ to ensure comparability with the SDSS data. This standardization may suppress fine-grained morphological details, especially for higher-resolution simulations. However, we mitigate potential biases via a $k$-sparse VAE, which focuses on robust, sparsity-induced features that are less sensitive to pixel-level artifacts. Additionally, we focussed on methodological advancements here and our method will equally be applicable to future higher resolution observational images.

\paragraph{Sample size}

We note that while our framework supports model comparison under limited simulation budgets, a minimal number of a few hundred simulation images per class is recommended to ensure classifier calibration and reliable posterior estimates. The term "sparse" in our title reflects both this constraint and our use of sparse latent representations in the VAE architecture.

Similarly, We include simulation variants of different resolutions (e.g., TNG50 vs. TNG100, NIHAO-UHD vs. NOAGN), and our classifier’s ability to partially separate these supports robustness to physical mass resolution changes. In fact, misclassification patterns (e.g., some confusion between TNG50 and TNG100) are consistent with the models’ intrinsic similarity and not indicative of methodological weaknesses.

\paragraph{Survey realism and OOD robustness:}

All simulation images undergo a consistent radiative transfer and noise-injection pipeline, designed to closely replicate SDSS observing conditions. This includes PSF convolution, noise injection, and color mapping, following the RealSim \citep{Bottrell_2019} procedures as discussed in section \ref{sec:methods}. Nonetheless, real survey effects such as background contamination (e.g., stars, cosmic rays) are not explicitly modeled. Our OOD detection strategy (via GEN scores) explicitly removes data that deviate in \textit{any} latent dimension---be it physical properties or observational artifacts---thus making our method agnostic to such nuisances. In practice, this ensures that only comparable (in-distribution) galaxies are evaluated, boosting the reliability of model comparison.

\paragraph{Model choice and representation learning:}

Our results rely on a stacking ensemble for classification and a $k$-sparse VAE for dimensionality reduction. Both choices are motivated by performance and interpretability. The ensemble method increases robustness under data scarcity, while the sparse VAE ensures disentanglement for physical interpretability (via SHAP values). While different model choices might yield slightly varied results, we have tested alternatives (e.g., different VAE architectures, varying VAE sparsity and loss functions) and found qualitative stability in conclusions. In particular, we extensively compared different classifiers to inspect the sensitivity of our result on the exact architecture. In all our experiments we have found our results to be qualitatively robust against any changes in the setup. Moreover, our explainable framework allows transparent inspection of learned features, enhancing interpretability and reproducibility.

\section{Summary and Conclusions}
\label{sec:conclusions}

In this paper we have set out to explore novel approaches to perform model comparison in the context of galaxy images and hydrodynamical simulations. To this extent we have developed novel methods for model misspecification detection and Bayesian model comparison.  By casting the Bayesian model comparison task as a classification task, we are able to select the relatively best matching model without the need for potentially lossy hand-crafted summary statistics. Furthermore our approach enables the usage of explainable AI techniques, such as SHAP values, to get a deeper insight into the advantages/disadvantages of individual models.
Additionally, our innovative approach for detecting model misspecification not only enables us to gauge the misfit of individual models, but also enables insights on why or in which respect these models are missing to model key physical aspects of galaxies. 

In detail, our model misspecification detection and Bayesian model comparison pipeline employs $k$-sparse VAEs as compression algorithms and performs model comparison based on classifier networks in the lower dimensional embedding space. Model misspecification detection is done via Generalized ENtropy scores before performing the model comparison. Our approach is especially well suited in caeses where simulation models are expensive to calculate and hence training data is scarce. With this paper, we not only derive these methods and their corresponding network architectures but we further provide thorough guidelines on how to calibrate classifiers to derive robust results.

Finally, we use these new methods to perform a quantitative comparison between various state-of-the-art hydrodynamical simulations taken from the IlustrisTNG project and the NIHAO project and gauge them against $\sim600,000$ real SDSS galaxy images.

Our results are summarized as follows:
\begin{itemize}
    \item We combine novel simulation-based Bayesian model comparison with a new misspecification detection technique to compare simulated galaxy images of 6 hydrodynamical models from the NIHAO and IllustrisTNG simulations against real observations from the Sloan Digital Sky Survey (SDSS). Thereby we address the typical problem of low simulation budgets by first training a $k$-sparse variational autoencoder (VAE) on the abundant observational dataset of SDSS images. The VAE learns to extract informative latent embeddings and delineates the typical set of real images. See \autoref{fig:pipeline} for a flow chart of our method. 
    \item Typically, hydrodynamical simulations might not explain every detail of real world galaxy images. To reveal these simulation gaps, we then perform out-of-distribution (OOD) detection based on the logit functions of classifiers trained on the embeddings of simulated images and find that all simulation models tested here are partially OOD (see e.g. \autoref{fig:umap} and \autoref{fig:gen-ood}). 
    \item We then perform amortized Bayesian model comparison using probabilistic classification only on the in-distribution data to ensure reliable comparison results. Our comparison approach is able to quantitatively identify the relatively best-performing model (see \autoref{fig:violin}) along with partial explanations through SHAP values (see \autoref{fig:shap}).
    \item We find that all 6 simulation models tested here are misspecified compared to real SDSS observations and can only explain part of reality. The relatively best performing model comes from the standard NIHAO simulations without Active Galactic Nuclei physics (see \autoref{fig:violin}). Carefully inspecting SHAP values we find that the main difference between NIHAO and IllustrisTNG is given by color and morphology -- NIHAO is redder and clumpier than IllustrisTNG (see \autoref{fig:shap}).
\end{itemize}

In conclusion, we developed and tested an innovative methods for simulation-based Bayesian model comparison with a new misspecification detection technique to compare simulated galaxy images produced by hydrodynamical models against real galaxy observations. By additionally using explainable AI methods such as SHAP values we are able to gain physical intuition about our results and help explaining which physical aspects of a particular simulation causes it to match real observations better or worse. This unique feature helps to inform simulators on how to improve their simulation model.

\section{Code Availability}
\label{sec:appendix_code}

We publicly release our code via Github: \url{https://github.com/z01ly/model-comparison}. For running our pipeline, see \texttt{src/main\_func.py} in the repository.

\begin{acknowledgements}
      This project was made possible by funding from the Carl Zeiss Stiftung.
\end{acknowledgements}

\bibliographystyle{aa}
\bibliography{bibtex/references}

\begin{appendix}

\section{Classifier calibration}
\label{sec:appendix_calibration}

\begin{figure*}
    \centering
    \subfigure[random forest]{
        \includegraphics[width=\hsize]{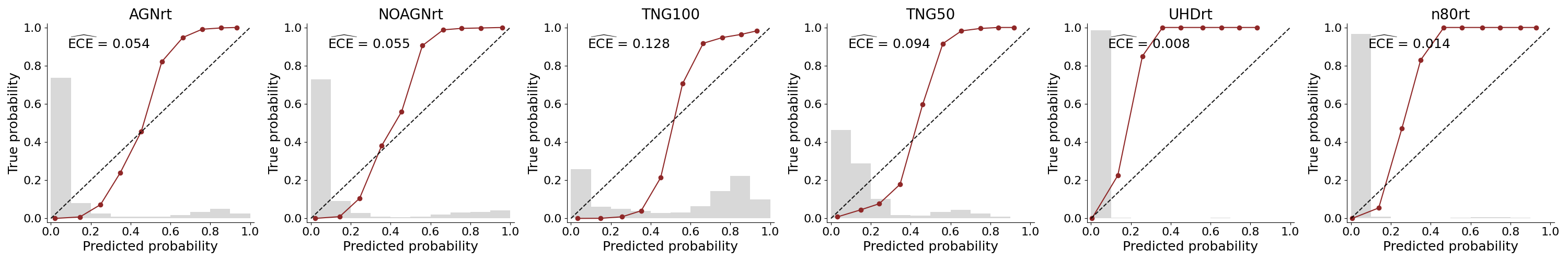}
    }
    \subfigure[XGBoost]{
        \includegraphics[width=\hsize]{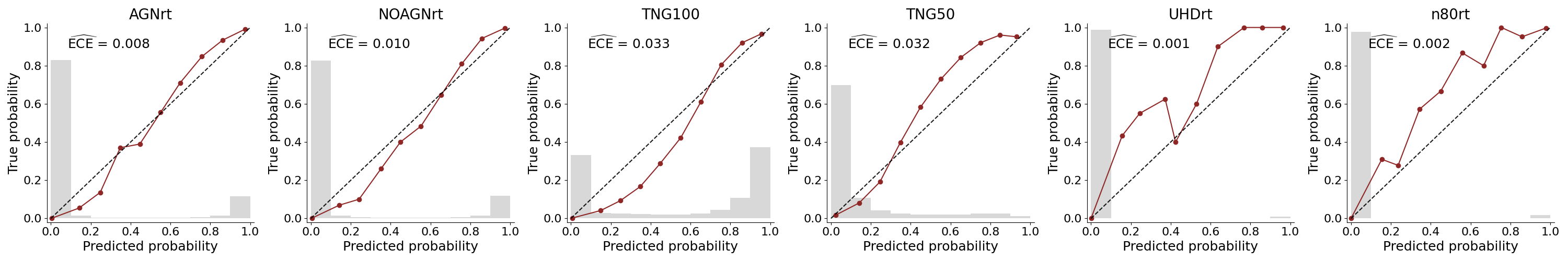}
    }
    \subfigure[stacking-MLP-RF-XGB]{
        \includegraphics[width=\hsize]{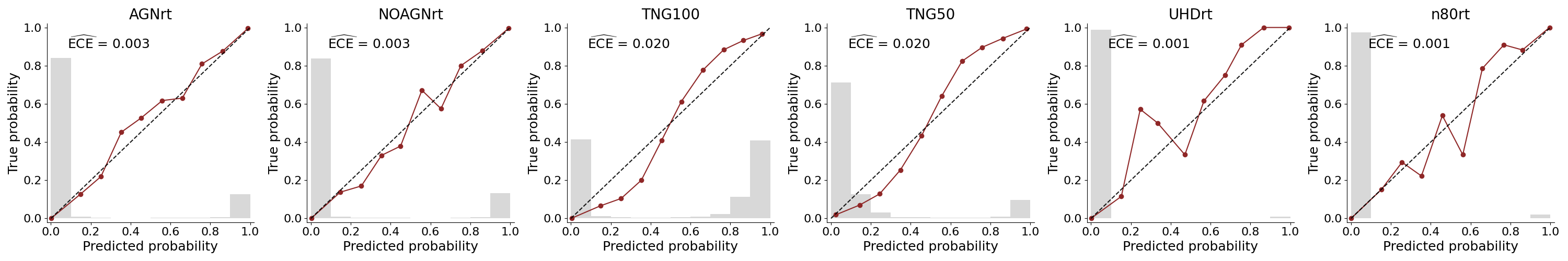}
    }
    \caption{Calibration curves of classifiers. Top: random forest, Middle: XGBoost, Bottom: stacking-MLP-RF-XGB}
    \label{fig:cc}
\end{figure*}

\begin{figure*}
    \centering
    \subfigure[random forest]{
        \includegraphics[width=.32\textwidth, trim={.55cm 0cm 3.cm 1.1cm}, clip]{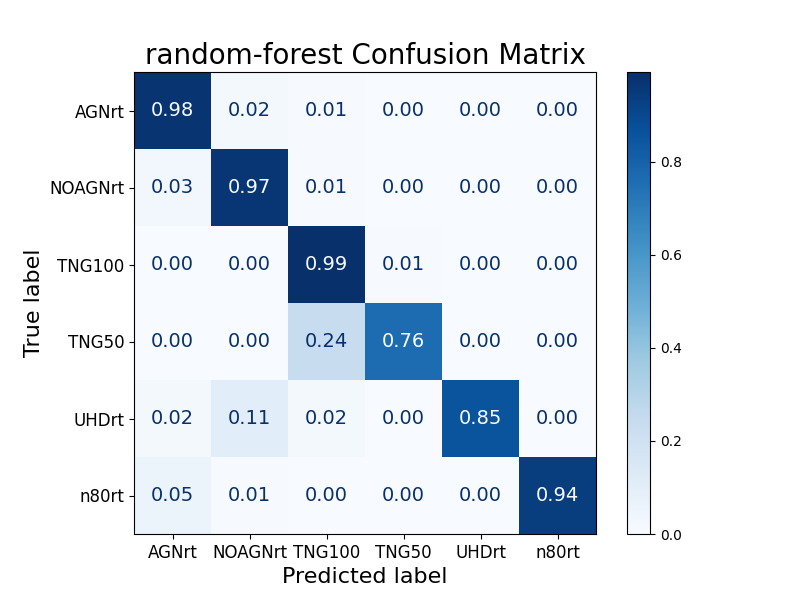}
    }
    \subfigure[XGBoost]{
        \includegraphics[width=.32\textwidth, trim={.55cm 0cm 3.cm 1.1cm}, clip]{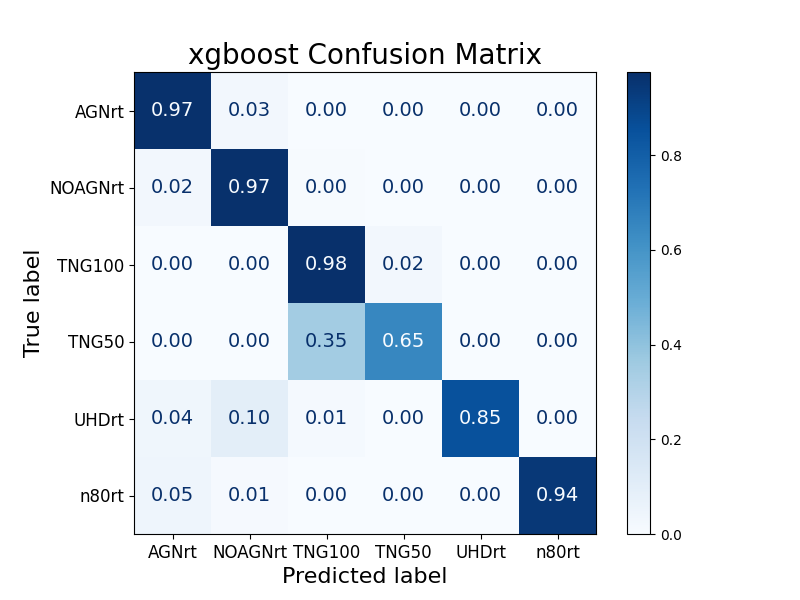}
    }
    \subfigure[stacking-MLP-RF-XGB]{
        \includegraphics[width=.32\textwidth, trim={.55cm 0cm 3.cm 1.1cm}, clip]{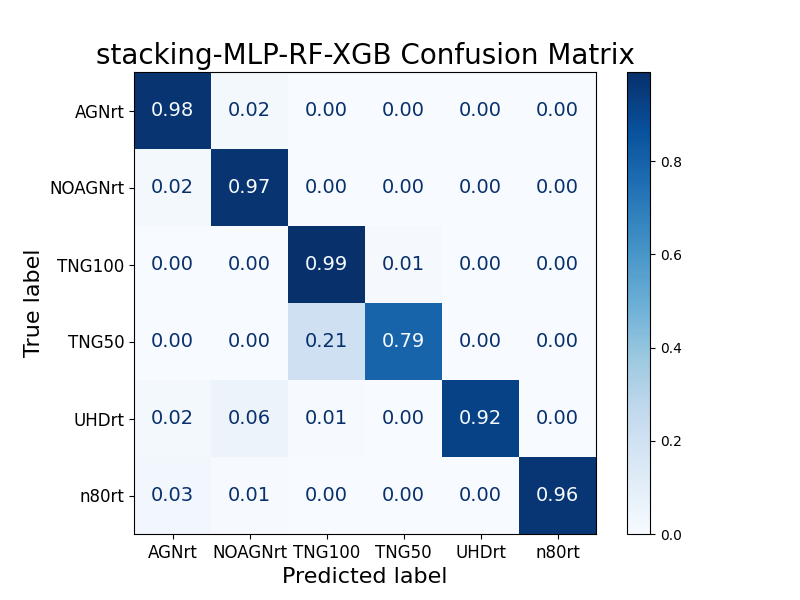}
    }
    \caption{Confusion matrices of classifiers. Top: random forest, Middle: XGBoost, Bottom: stacking-MLP-RF-XGB}
    \label{fig:cm}
\end{figure*}

\section{Sensitivity analysis of OOD threshold}
\label{subsec:ood_threshold_analysis}

Our main results are based on setting an OOD threshold of 95\% in the reference GEN score distribution generated from the simulation test set. Here we adjust the threshold (99\%, 97\%, 95\%, 93\% 90\% and 85\%) to explore how the model comparison results and hence the violin plots and the percent of out-of-distribution SDSS test data changes accordingly. 

\begin{figure*}[h]
    \centering
    \subfigure[threshold vs. ID ratio]{
        \includegraphics[width=0.45\hsize]{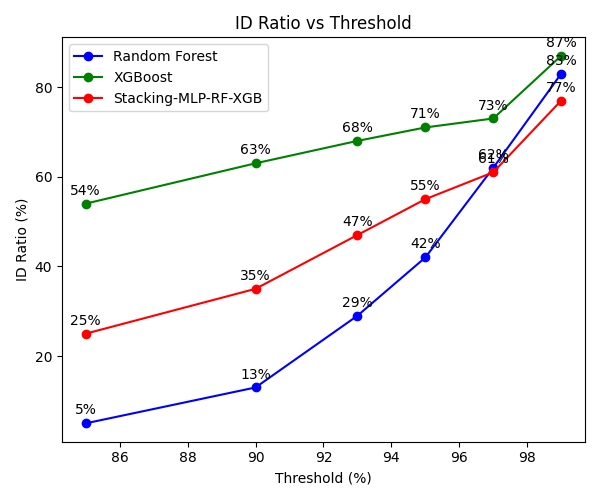}
    }
    \subfigure[threshold vs. OOD ratio]{
        \includegraphics[width=0.45\hsize]{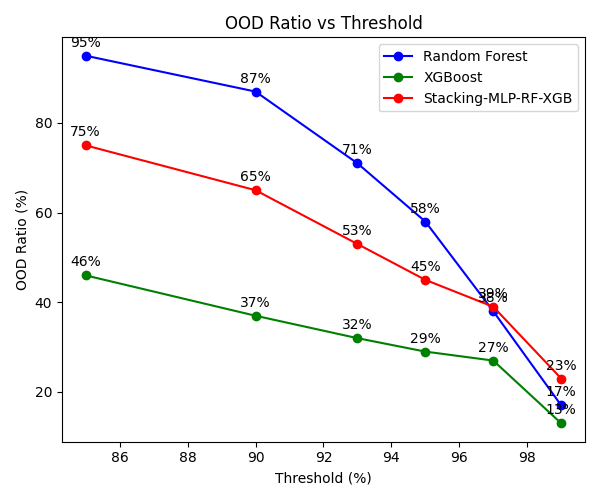}
    }
    \caption{The relationship between threshold choice and ID / OOD ratio of random forest, XGBoost and stacking-MLP-RF-XGB.}
    \label{fig:threshold_ratio}
\end{figure*}

We illustrate the relationship between threshold value and ID / OOD ratio of the random forest, XGBoost and stacking-MLP-RF-XGB classifiers in \autoref{fig:threshold_ratio}. The curve of random forest is the steepest one, showing that it is sensitive to the threshold change. As the threshold increases, ID / OOD ratios from the 3 ensembles tend to be more similar. 

\begin{figure*}
    \centering
    \subfigure[random forest]{
        \includegraphics[width=0.32\hsize, trim={1.5cm 0cm 2.5cm 1.cm}, clip]{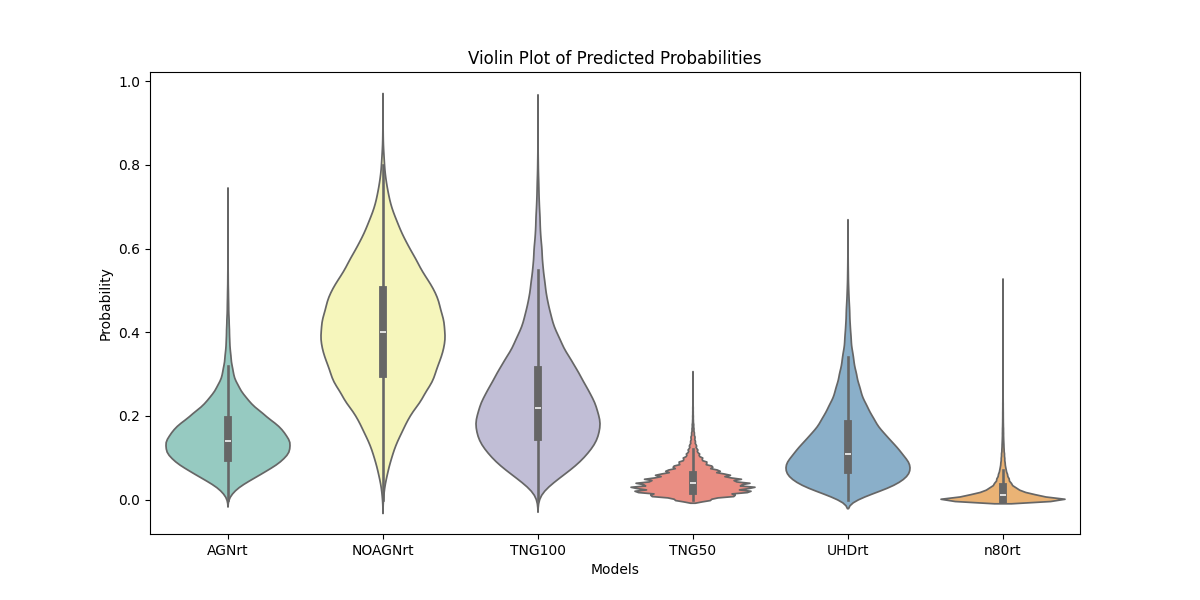}
    }
    \subfigure[XGBoost]{
        \includegraphics[width=0.32\hsize, trim={1.5cm 0cm 2.5cm 1.cm}, clip]{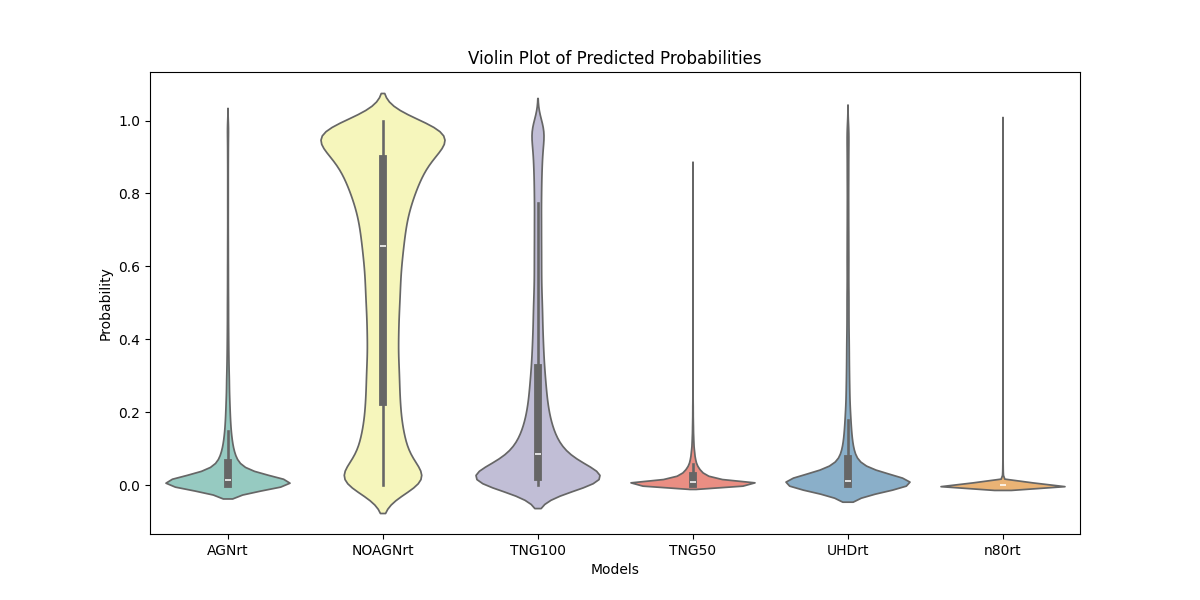}
    }
    \subfigure[stacking-MLP-RF-XGB]{
        \includegraphics[width=0.32\hsize, trim={1.5cm 0cm 2.5cm 1.cm}, clip]{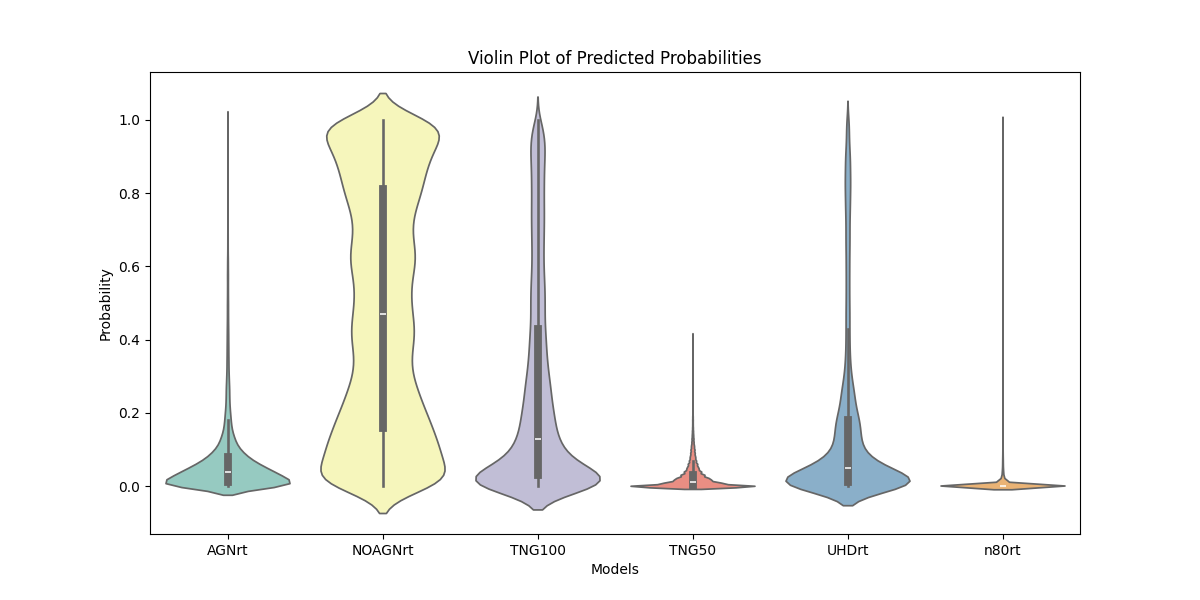}
    }
    \caption{Violin plots from classifying the whole SDSS test set. Left: random forest, Middle: XGBoost, Right: stacking-MLP-RF-XGB}
    \label{fig:violin-sdss-all}
\end{figure*}

To present the relationship between threshold and our final violin plot, we use the violin plots from classifying the whole SDSS test set as a reference, see \autoref{fig:violin-sdss-all}. The violin plots corresponding to threshold values of 99\%, 97\%, 95\%, 93\%, 90\% and 85\% are shown in \autoref{fig:violin-percent99}, \autoref{fig:violin-percent97}, \autoref{fig:violin},\autoref{fig:violin-percent93}, \autoref{fig:violin-percent90} and \autoref{fig:violin-percent85}, respectively. Compared to the reference violin plots, discarded out-of-distribution SDSS test data are mainly those classified as 40\% to 60\% NOAGN or TNG100. This is reasonable since a probability around 50\% implies that the classifier does not know how to handle these data. This in turn confirms OOD. We take a look at two extreme cases: for threshold 99\%, to little SDSS data is discarded, while for threshold 85\%, too much data are discarded as the discarded range of NOAGN has expanded from 20\% to 80\%. 

Hence, we empirically decide for a threshold value by considering both the line chart \autoref{fig:threshold_ratio} and the violin plots from different thresholds. We find that a threshold between $\sim$93\% and $\sim$97\% is reasonable -- justifying our threshold choice of 95\% in the main pipeline.

\foreach \i in {99, 97, 93, 90, 85}{
\begin{figure*}
    \centering
    \subfigure[random forest]{
        \includegraphics[width=0.32\hsize, trim={1.5cm 0cm 2.5cm 1.cm}, clip]{figs/gen-ood/re-classify/percent\i/violin-plot/random-forest-violin.png}
    }
    \subfigure[XGBoost]{
        \includegraphics[width=0.32\hsize, trim={1.5cm 0cm 2.5cm 1.cm}, clip]{figs/gen-ood/re-classify/percent\i/violin-plot/xgboost-violin.png}
    }
    \subfigure[stacking-MLP-RF-XGB]{
        \includegraphics[width=0.32\hsize, trim={1.5cm 0cm 2.5cm 1.cm}, clip]{figs/gen-ood/re-classify/percent\i/violin-plot/stacking-MLP-RF-XGB-violin.png}
    }
    \caption{Violin plots of threshold \i\%. Left: random forest, Middle: XGBoost, Right: stacking-MLP-RF-XGB}
    \label{fig:violin-percent\i}
\end{figure*}
}

\section{SHAP plots of AGN, TNG50, UHD and n80}
\label{sec:appendix_shap_plots}

Here we present SHAP plots of AGN, TNG50, UHD and n80 from XGBoost classifier. 

\begin{figure*}[h]
    \centering
    \subfigure[AGN]{
        \includegraphics[width=0.8\hsize]{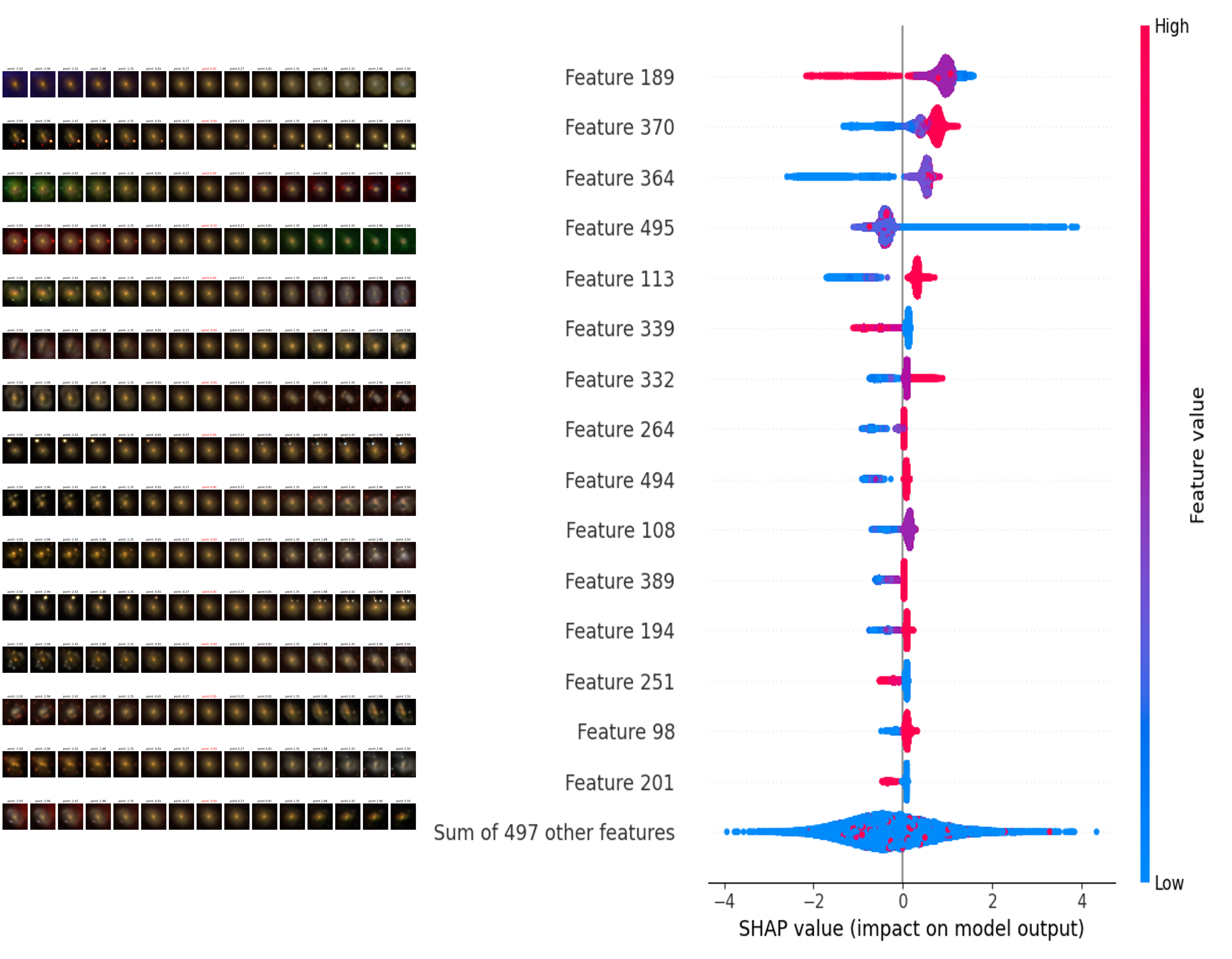}
    }
    \subfigure[TNG50]{
        \includegraphics[width=0.8\hsize]{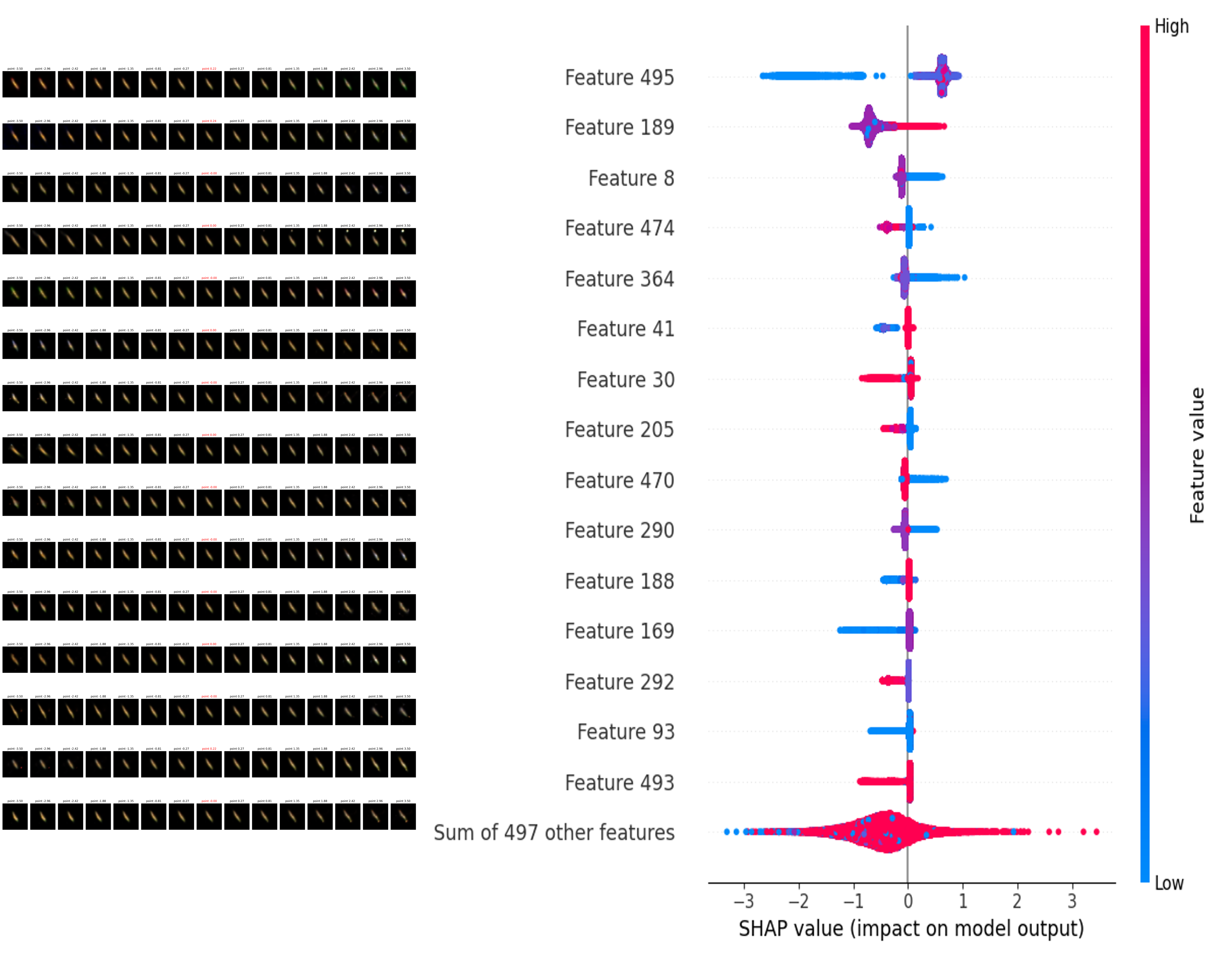}
    }
    \caption{Same as \autoref{fig:shap} but for AGN and TNG50}
    \label{fig:shap-appendix-1}
\end{figure*}

\begin{figure*}[h]
    \centering
    \subfigure[UHD]{
        \includegraphics[width=0.8\hsize]{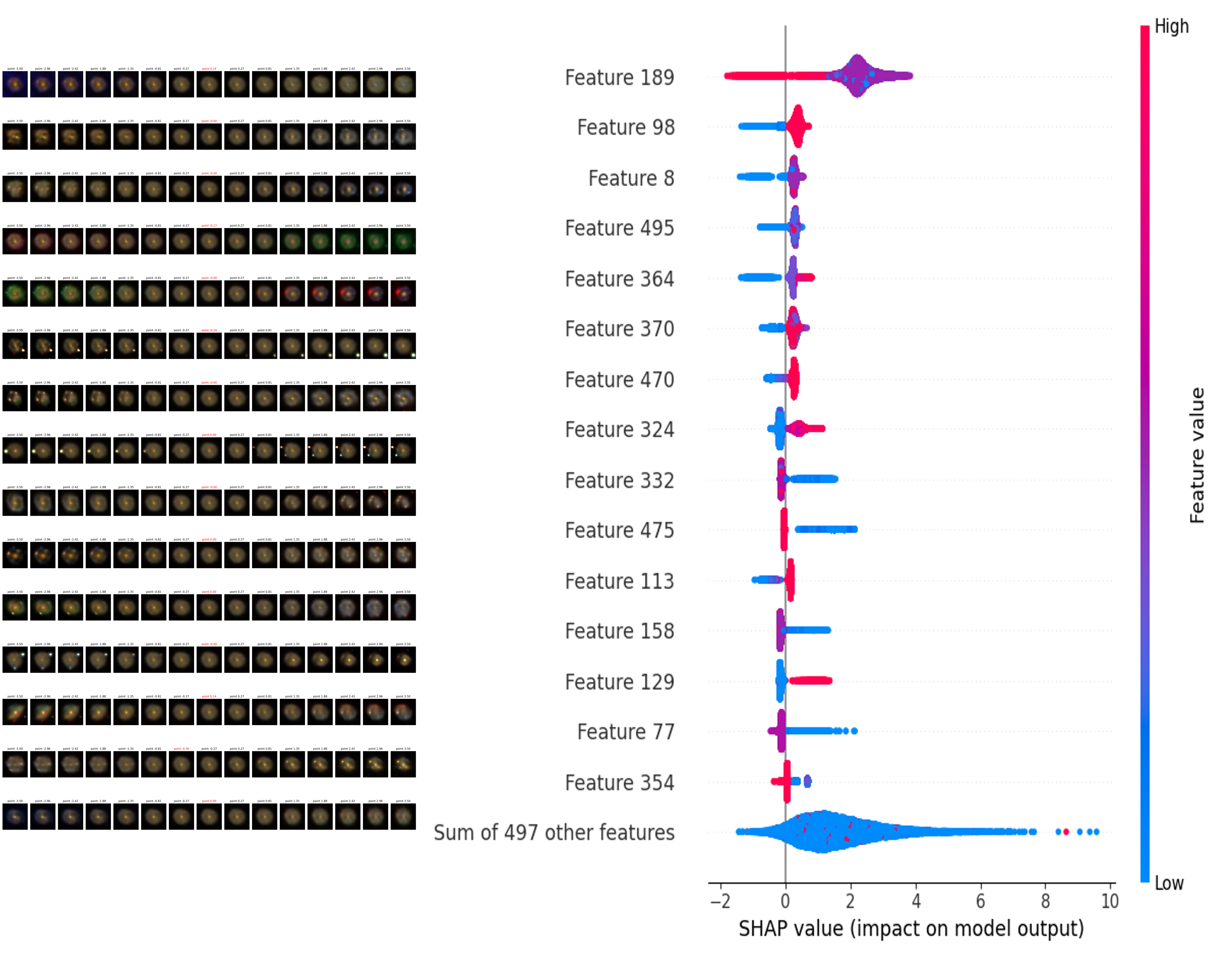}
    }
    \subfigure[n80]{
        \includegraphics[width=0.8\hsize]{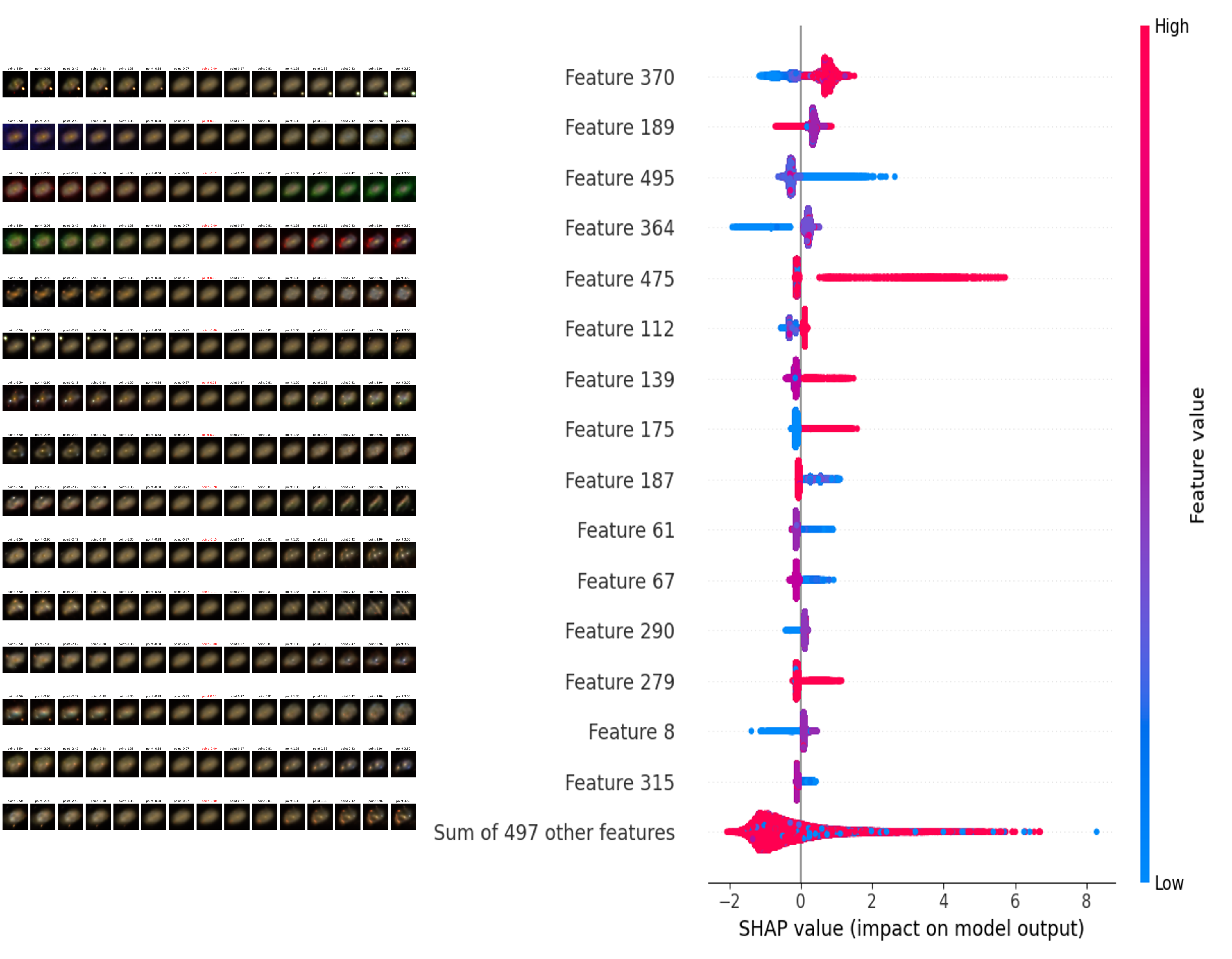}
    }
    \caption{Same as \autoref{fig:shap} but for UHD and n80}
    \label{fig:shap-appendix-2}
\end{figure*}

\end{appendix}

\end{document}